\documentclass[prc]{revtex4}

\usepackage{graphicx}
\usepackage{amssymb}
\usepackage{amsmath}
\usepackage{dsfont}
\usepackage[colorlinks=true]{hyperref}
\usepackage{multirow}

\newcommand{\be}{\begin{equation}}
\newcommand{\ee}{\end{equation}}
\newcommand{\ba}{\begin{eqnarray}}
\newcommand{\ea}{\end{eqnarray}}
\newcommand{\ban}{\begin{eqnarray*}}
\newcommand{\ean}{\end{eqnarray*}}
\newcommand \nn {\nonumber}

\begin{document}

\title{Heavy Quarks in Turbulent QCD Plasmas}

\author{Stanis\l aw Mr\' owczy\' nski}

\affiliation{Institute of Physics, Jan Kochanowski University, ul. \'Swi\c etokrzyska 15, PL-25-406 Kielce, Poland 
\\
and National Centre for Nuclear Research, ul. Ho\. za 69, PL-00-681 Warsaw, Poland}

\date{October 23, 2017}

\begin{abstract}

The quark-gluon plasma, which is produced at an early stage of ultrarelativistic heavy-ion collisions, is expected to be initially strongly populated with chromodynamic fields. We address the question how heavy quarks interact with such a turbulent plasma in comparison with an equilibrated one of the same energy density. For this purpose we derive a Fokker-Planck transport equation of heavy quarks embedded in a plasma of light quarks and gluons. We first discuss the equilibrium plasma and then the turbulent one  applying the same approach, where the heavy quarks interact not with the plasma constituents but rather with the long wavelength classical fields. We first consider the three schematic models of isotropic trubulent plasma and then the simplified model of glasma with the chromodynamic fields only along the beam direction. The momentum broadening and collisional energy loss of a test heavy quark are computed and compared to those of equilibrium plasma of the same energy density.

\end{abstract}

\maketitle

\section{Introduction}

The early stage of relativistic heavy-ion collisions is least known because there are hardly any experimentally accessible signals of the phase. Nevertheless one expects that the quark-gluon plasma, which is produced in the collisions, is initially strongly populated with chromodynamic fields. Within the framework of the Color Glass Condensate (CGC) approach, see {\it e.g.} the review \cite{Gelis:2012ri}, color charges of partons confined in the colliding nuclei act as sources of long wavelength chromodynamic fields which can be treated classically because of large occupation numbers of the soft modes. Since the density of color charges per transverse area of heavy nuclei is large, the corresponding momentum scale $Q_s$ is expected to be significantly bigger than the QCD scale parameter $\Lambda_{\rm QCD}$. Consequently, the coupling constant $\alpha_s$ is presumably sufficiently small and perturbative methods are applicable. The system, however, is rather strongly interacting because of the high-amplitude fields present in the system. 

A momentum anisotropy of the early stage quark-gluon plasma makes it unstable with respect to chromomagnetic modes which in turn cause a spontaneous generation of the fields, as explained at length in the review article \cite{Mrowczynski:2016etf}. Therefore, the effect of strong fields is further enhanced. Following the terminology of electromagnetic plasma, we call such a nonequilibrium system of fields as the {\it turbulent} plasma meaning that numerous modes are excited in the system. In the CGC approach the non-equilibrium system of fields from the early stage of relativistic heavy-ion collisions is called {\it glasma} \cite{Gelis:2012ri} and it can be treated as a specific realization of the turbulent QCD plasma. Leaving aside the mechanism of field generation and its detailed structure, one asks what are the transport properties of turbulent plasmas. We are specifically interested how heavy quarks - charm or beauty - behave in such a system when compared to the equilibrium plasma of the same energy density. 

Heavy quarks are often treated as a probe of strongly interacting matter created in relativistic heavy-ion collisions, see e.g. the review \cite{Prino:2016cni}. Thanks to their large masses the quarks are produced only at the earliest stage of the collision due to hard interactions of partons from incoming nuclei. Later on they propagate through a surrounding medium testing the entire history of the system. It has been long believed that the interaction of heavy quarks is significantly weaker than that of light quarks or gluons but experimental data clearly contradict the expectation. As discussed in the review \cite{Prino:2016cni}, the behavior of mesons containing a heavy quark is rather similar to that of light mesons at both small and large transverse momenta. The problem is not fully resolved. 

The medium created in relativistic heavy-ion collisions evolves fast towards the locally equilibrated quark-gluon plasma which expands hydrodynamically and ultimately is converted into a hadron gas. Final momentum spectra of heavy quarks are mostly shaped in the long-lasting equilibrium phase which is relatively well understood. An effect of a pre-equilibrium phase is often entirely ignored but this transient phase can significantly influence heavy-quark spectra because of its high density. Non-equilibrium calculations recently performed in a framework of kinetic theory \cite{Das:2017dsh} confirm the suggestion. However, we are interested even in the earlier phase when the medium is not described in terms of quasi-particles, as in a kinetic theory, but rather as a system dominated by classical fields which is the turbulent plasma. 

A simple parametric estimate suggests that the interaction of heavy quarks in a turbulent plasma is much stronger than in the equilibrium one, if the coupling constant $g$ is small. The momentum broadening parameter $\hat{q}$, for example, is of order $g^4$ in equilibrium plasmas. Since the quark of interest actually interacts with soft gluons emitted by plasma constituents, one can think that the factor $g^4$ is composed of two pieces of $g^2$. The first one is related to the gluon emission and the second one to the gluon absorption. If soft chromodynamic fields are present in the plasma, the interaction of the quark should be rather of order $g^2$ than $g^4$. In Sec.~\ref{sec-eq-vs-turb} we argue that $\hat{q}$ is indeed not of the order $g^4$, not even $g^2$ but presumably of the order $g$ in a turbulent plasma. 

Because of their big masses, relaxation times of heavy quarks, which are produced in relativistic heavy-ion collisions, are expected to be significantly longer than that of light quarks and gluons. When an equilibrium or, more generally, a stationary state is reached by light quarks and gluons, heavy quarks need some extra time to adjust to the state of the plasma. Such a situation is naturally described in terms of the Fokker-Planck transport equation which was indeed repeatedly applied to heavy quarks in  \cite{Moore:2004tg,Svetitsky:1987gq,vanHees:2004gq,Mustafa:2004dr}. The equation is usually derived from the Boltzmann equation by applying the so-called diffusion approximation to the collision term \cite{LP81}. The approximation assumes that the momentum transfer to the heavy quark in every collision is much smaller than the quark momentum.

The aim of this paper is threefold. In the first part we rederive the Fokker-Planck equation of heavy quarks which do not interact with plasma constituents but rather with soft classical fields present in the plasma. Specifically, we apply the so-called quasi-linear theory known from the electromagnetic plasma  \cite{Ved61,LP81}. The theory assumes that the distribution function can be decomposed into a large but slowly varying regular part and a small fluctuating or turbulent one which oscillates fast. The average over a statistical ensemble of the turbulent part is assumed to vanish and thus the average of the distribution function equals its regular part. The turbulent contribution to the distribution function obeys the collisionless transport equation while the transport equation of the regular part is determined by the fluctuation spectra which provide the collision term. The derivation presented here closely follows the procedure which was developed for QCD in \cite{Mrowczynski:2009gf}, where, however, only the longitudinal chromoelectric field was taken into account and here the complete chromodynamic field is considered. The equilibrium correlation functions of chromodynamic fields, which are needed to obtain the quasi-linear transport equations, were derived in \cite{Mrowczynski:2008ae}. 

Our second aim is to confront the equilibrium plasma with the turbulent one. Therefore, we consider three models of isotropic turbulent plasma in the second part of the paper. Postulating a form of the correlation functions of chromodynamic fields, we derive the coefficients of the Fokker-Planck equation which can be related to the energy loss, momentum broadening and diffusion coefficient of heavy quarks in the plasma. The transport coefficients of turbulent plasma are compared to those of the equilibrium one at the same energy density.

The third aim is to study an evolution of heavy quarks at the earliest stage of relativistic have-ion collisions. Since the chromoelectric and chromomagnetic fields spanned between the receding nuclei are initially mostly parallel to the beam direction, we model the glasma with the boost invariant correlation functions of longitudinal fields. The energy loss and momentum broadening of heavy quarks are computed, assuming that all energy of the glasma is accumulated in the longitudinal fields.  

At the end of the introductory remarks we note that an approach similar to ours, which was also inspired by the electromagnetic plasma studies \cite{Dupree:1966}, was formulated in \cite{Asakawa:2006jn}, see  \cite{Majumder:2007zh,Chandra:2015gma} as well. We also mention an attempt  \cite{Song:2015jmn} to study transport of heavy quarks in a plasma populated by strong chromodynamic fields. Unfortunately, the paper is flawed as the framework of an isotropic Langevin approach is applied to anisotropic plasmas.

Throughout the paper we use the natural system of units with $c = \hbar = k_B =1$; our choice of the signature of the metric tensor is $(+ - - -)$. Lorentz indices are denoted with $\mu, \, \nu = 0, \, 1, \, 2, \,3$ and $i,\, j = 1, \, 2, \,3$ label the Cartesian coordinates $x,\, y, \, z$. The color indices of the adjoint representation of ${\rm SU}(N_c)$ gauge group are $a,\, b = 1, \, 2, \dots N_c^2 -1$.

\section{Derivation of Fokker-Planck equation}
\label{sec-derivation}

Our derivation of the Fokker-Planck equation of heavy quarks embedded in quark-gluon plasma starts with the transport equation of the Vlasov form
\be
\label{transport-eq}
\big(D^0 +{\bf v} \cdot {\bf D}\big)Q(t,{\bf r},{\bf p})
- {1 \over 2}
\big{\{}{\bf F}(t,{\bf r}) , \nabla_p Q(t,{\bf r},{\bf p}) \big{\}} = 0 ,
\ee
where the distribution function $Q(t,{\bf r},{\bf p})$ of heavy quarks is the $N_c\times N_c$ hermitian matrix which belongs to the fundamental representation of the SU($N_c$) group. The distribution function depends on the time ($t$), position (${\bf r}$) and momentum (${\bf p}$) variables. There is no explicit dependence on the time-like component of the four-momentum $p^\mu = (p^0, {\bf p})$ as the distribution function is assumed to be non-zero only for momenta obeying the mass-shell constraint that is $p_0 = E_{\bf p}=\sqrt{{\bf p}^2 + m^2}$. The quark velocity equals ${\bf v}={\bf p}/E_{\bf p}$ and  $D^\mu \equiv (D^0, {\bf D}) \equiv \partial^\mu - ig [A^\mu(x),\cdots \;]$ with $A^\mu(x)$ being the chromodynamic potential in the fundamental representation. The mean-field term of the transport equation (\ref{transport-eq}) is expressed through the color Lorentz force ${\bf F}(t,{\bf r}) \equiv g \big({\bf E}(t,{\bf r}) + {\bf v} \times {\bf B}(t,{\bf r})\big)$ with the chromoelectric ${\bf E}(t,{\bf r})$ and chromomagnetic ${\bf B}(t,{\bf r})$ fields also belonging to the fundamental representation. The symbol $\{\dots , \dots \}$ denotes the anticommutator. The derivation of the transport equation (\ref{transport-eq}) is discussed in detail in the review \cite{Mrowczynski:2016etf}.

Further on we assume that the chromodynamic fields and the distribution function which enter the transport equation (\ref{transport-eq}) can be decomposed into a {\it regular} and {\it fluctuating} or {\it turbulent} component. The distribution function is thus written down as
\be
\label{reg-fluc}
Q(t,{\bf r},{\bf p}) = \langle Q(t,{\bf r},{\bf p}) \rangle 
+ \delta Q(t,{\bf r},{\bf p}) ,
\ee
where $\langle \cdots \rangle$ denotes ensemble average; $\langle Q(t,{\bf r},{\bf p}) \rangle$ is called the regular part while $\delta Q(t,{\bf r},{\bf p})$ is called the fluctuating or turbulent one. It directly follows from Eq.~(\ref{reg-fluc}) that $\langle \delta Q \rangle =0$. The regular contribution is assumed to be color neutral or white, and it is expressed as   
\be
\label{whiteness}
\langle Q(t,{\bf r},{\bf p}) \rangle  = 
n (t,{\bf r},{\bf p}) \, \mathds{1}, 
\ee
where $\mathds{1}$ is the unit matrix in color space. Since the distribution function transforms under gauge transformations as $Q \rightarrow U\, Q U^\dagger$, where $U$ is the transformation matrix, the regular contribution of the form (\ref{whiteness}) is gauge independent. We also assume that
\be
\label{quasi-linear-conditions-1}
|\langle Q \rangle | \gg |\delta Q| \;, \qquad
|\nabla_p \langle Q \rangle | \gg | \nabla_p \delta Q| ,
\ee
but at the same time 
\be
\label{quasi-linear-conditions-2}
\left|\frac{\partial \delta Q }{\partial t} \right| 
\gg \left|\frac{\partial \langle Q \rangle}{\partial t} \right| 
\;, \qquad
| \nabla \delta Q| \gg |\nabla \langle Q \rangle | .
\ee
What concerns the chromodynamic fields, we assume in accordance with Eq.~(\ref{whiteness}) that their regular parts vanish and thus 
\be
\langle {\bf E}(t,{\bf r}) \rangle = \langle {\bf B}(t,{\bf r}) \rangle = 0 .
\ee

We substitute the distribution function (\ref{reg-fluc}) into the transport equation (\ref{transport-eq}) and linearize the equations in the fluctuating contributions. Thus we get the equation
\be
\label{trans-eq-lin}
{\cal D}\, \delta Q(t,{\bf r},{\bf p}) -  {\bf F}(t,{\bf r}) \cdot \nabla_p n(t,{\bf r},{\bf p}) = 0 , 
\ee
where ${\cal D} \equiv  \frac{\partial}{\partial t} + {\bf v}\cdot \nabla $ is the substantial or material derivative. 

Now we substitute the distribution functions (\ref{reg-fluc}) into the transport equations (\ref{transport-eq}) but instead of linearizing the equation in the fluctuating contributions, we take the ensemble average of the resulting equation and trace over the color indices. Thus we get
\be
\label{trans-eq-reg}
{\cal D}\, n(t,{\bf r},{\bf p})  - \frac{1}{N_c}{\rm Tr} 
\big\langle {\bf F}(t,{\bf r}) \cdot \nabla_p \delta Q(t,{\bf r},{\bf p})  \big\rangle 
= 0 .
\ee
Since the regular part of distribution function is assumed to be color neutral, see Eq.~(\ref{whiteness}), the term ${\rm Tr}[\langle {\bf F} \cdot \nabla_p n \rangle]$ vanishes because the fields ${\bf E}, \; {\bf B}$ are traceless. The trace over color indices also cancels the terms originating from covariant derivatives like ${\rm Tr}\langle [A^\mu,\delta Q] \rangle$. We finally note that the trace ${\rm Tr}[\langle {\bf F} \cdot \nabla_p \delta Q \rangle]$ is gauge independent as the regular distribution function $n(t,{\bf r},{\bf p})$ is.

Now, we are going to write down the transport equation (\ref{trans-eq-reg}) in the Fokker-Planck form. For this purpose we observe that due to the condition (\ref{quasi-linear-conditions-2}), the space-time dependence of the regular distribution function can be neglected in the linearized transport equation (\ref{trans-eq-lin}) and then, the equation becomes easily solvable. We solve it with the initial condition
\be
\delta Q(t=0,{\bf r},{\bf p}) = \delta Q_0({\bf r},{\bf p}) ,
\ee
using the one-sided Fourier transformation defined as
\be
\label{1sF}
f(\omega,{\bf k}) = \int_0^\infty dt \int d^3r 
e^{i(\omega t - {\bf k}\cdot {\bf r})} f(t,{\bf r}) .
\ee
The inverse transformation is 
\be
\label{inv-1sF}
f(t,{\bf r}) = \int_{-\infty +i\sigma}^{\infty +i\sigma}
{d\omega \over 2\pi} \int {d^3k \over (2\pi)^3} 
e^{-i(\omega t - {\bf k}\cdot {\bf r})} f(\omega,{\bf k}) ,
\ee
where the real parameter $\sigma > 0$ is chosen in such a way that the integral over $\omega$ is taken along a straight line in the complex $\omega-$plane, parallel to the real axis,  above all singularities of $f(\omega,{\bf k})$. 

The linearized transport equation (\ref{trans-eq-lin}), which is converted into the algebraic equation by means of the one-sided Fourier transformation, is
solved as
\be
\label{solution1}
\delta Q(\omega,{\bf k},{\bf p}) = i\frac{{\bf F}(\omega,{\bf k}) \cdot \nabla_p n({\bf p}) +  \delta Q_0({\bf k},{\bf p})}
{\omega - {\bf k}\cdot {\bf v}} ,
\ee
We stress that although we have ignored the (weak) frequency and wave number dependence of the regular distribution $n$, the fields ${\bf E}(\omega,{\bf k}),\; {\bf B}(\omega,{\bf k})$ retain their full frequency and wave number dependence in the expression (\ref{solution1}). Inverting the one-sided Fourier transformation, one finds the solution of the linearized transport equation as
\be
\label{sol-x}
\delta Q(t,{\bf r},{\bf p}) =
\int_0^t dt' \: {\bf F}\big(t', {\bf r}-{\bf v} (t-t')\big) \cdot \nabla_p n({\bf p})
+ \delta Q_0({\bf r}-{\bf v} t,{\bf p}) ,
\ee
where we assumed that ${\bf E}(\omega,{\bf k})$ and ${\bf B}(\omega,{\bf k})$ are analytic functions of $\omega$. 

With the help of the solution (\ref{sol-x}), the force term in the transport equation (\ref{trans-eq-reg}) becomes 
\ba
\label{E-nabla-dQ}
\big\langle {\bf F}(t,{\bf r}) \cdot \nabla_p \delta Q (t,{\bf r},{\bf p}) \big\rangle 
= \int_0^t dt' \: \nabla_p^i \big\langle F^i(t, {\bf r}) \, F^j\big(t', {\bf r}-{\bf v} (t-t')\big) \big\rangle \nabla_p^j n({\bf p}) 
+ \nabla_p^i \big\langle F^i(t, {\bf r}) 
\delta Q_0({\bf r}-{\bf v} t,{\bf p})\big\rangle .
\ea
The second term in the r.h.s. of Eq.~(\ref{E-nabla-dQ}) can be manipulated to the form
\be
\label{Y-def}
\frac{1}{N_c} {\rm Tr} \big[ \big\langle F^i(t, {\bf r}) 
\delta Q_0({\bf r}-{\bf v} t,{\bf p})\big\rangle \big] = Y^i({\bf v}) \, n({\bf p}) ,
\ee
which is effectively the definition of the vector $Y^i({\bf v})$. We also introduce the tensor
\be
\label{X-def}
X^{ij}({\bf v}) \equiv
\frac{1}{N_c} \int_0^t dt' \: {\rm Tr}\big[ \big\langle F^i(t, {\bf r}) F^j\big(t', {\bf r}-{\bf v} (t-t')\big) \big\rangle \big] ,
\ee
and we note that, as explained in the subsequent sections, $X$ and $Y$ become time independent for a sufficiently long $t$. Then, the transport equation (\ref{trans-eq-reg}) can be written as the Fokker-Planck equation
\be
\label{F-K-eq}
\Big({\cal D} - \nabla_p^i  X^{ij}({\bf v}) \nabla_p^j - \nabla_p^i  Y^i({\bf v}) \Big) n(t, {\bf r},{\bf p}) = 0.
\ee

Since the distribution function $n(t, {\bf r},{\bf p})$ carries no information about color degrees of freedom, the function is gauge invariant, and consequently $X^{ij}({\bf v})$ and $Y^i({\bf v})$ should be gauge invariant as well. However, one observes that $X^{ij}({\bf v})$ and $Y^i({\bf v})$ as defined by Eqs.~(\ref{Y-def}) and (\ref{X-def}) are gauge dependent because the traces are of nonlocal quantities in the definitions (\ref{Y-def}) and (\ref{X-def}). The starting transport equation (\ref{transport-eq}) is gauge covariant but the linearization procedure breaks the covariance because the covariant derivative is replaced by the normal one. Consequently, the solution (\ref{sol-x}) is not gauge covariant -- the right-hand side of Eq.~(\ref{sol-x}) transforms differently under local gauge transformations than the left-hand side. To cure the problem, one modifies the solution (\ref{sol-x}) by means of the link operator which is also called the gauge parallel transporter, see {\it e.g.} Sec.~IIIE of the review article \cite{Mrowczynski:2016etf}. Then, the modified solution obeys Eq.~(\ref{trans-eq-reg}) with the covariant derivative instead of the normal one. Let us briefly discuss the procedure in a context of the Fokker-Planck equation (\ref{F-K-eq}). 

According to Eq.~(\ref{X-def}), the tensor $X^{ij}({\bf v})$ is determined by the traces of the field correlation functions like $\langle E^i_a(t_1,{\bf r}_1)\, E^j_a(t_2,{\bf r}_2) \rangle$ where chromodynamic fields are written in the adjoint representation of the ${\rm SU}(N_c)$ group which is used further on. The trace becomes gauge invariant under the replacement 
\be
\langle E^i_a(t_1,{\bf r}_1)\, E^j_a(t_2,{\bf r}_2) \rangle \longrightarrow
\langle E^i_a(t_1,{\bf r}_1)\, \Omega_{ab}(t_1,{\bf r}_1|t_2,{\bf r}_2) \, E^j_b(t_2,{\bf r}_2) \rangle ,
\ee
where $\Omega_{ab}(t_1,{\bf r}_1|t_2,{\bf r}_2) $ is the link operator defined as 
\be
\label{link-def-fund}
\Omega (t_1,{\bf r}_1|t_2,{\bf r}_2) = {\cal P} \exp\Big[ ig \int_{(t_2,{\bf r}_2)}^{(t_1,{\bf r}_1)} ds_\mu A^\mu_c(s) T^c\Big] .
\ee
Here $T^c$ is the adjoint representation generator of the ${\rm SU}(N_c)$ group and ${\cal P}$ denotes the ordering along the path connecting the points $(t_2,{\bf r}_2)$ and $(t_1,{\bf r}_1)$. Since the fields transform as vectors under the local gauge transformation ${\cal U}(t,{\bf r})$ and the link transforms as 
\ba
\label{field-link-trans}
\Omega(t_1,{\bf r}_1|t_2,{\bf r}_2) \longrightarrow {\cal U} (t_1,{\bf r}_1) \; \Omega(t_1,{\bf r}_1|t_2,{\bf r}_2)\; {\cal U}^T(t_2,{\bf r}_2) ,
\ea
one checks that the trace of the correlation function which includes the link is indeed gauge invariant. Consequently, the tensor $X^{ij}({\bf v})$ is gauge invariant. Analogously one achieves the gauge invariance of the vector $Y^i({\bf v})$. Further on, whenever any nonlocal correlation function shows up a presence of the link operator is implicitly assumed even so it is not explicitly written.

\section{Fokker-Planck equation}
\label{sec-FP-eq}

Although this is a textbook material we briefly discuss here the Fokker-Planck equation (\ref{F-K-eq}). We first note that in the isotropic plasma the tensor $X^{ij}({\bf v})$ and vector $Y^i({\bf v})$ both depend on a single vector that is the heavy-quark velocity ${\bf v}$. Therefore, they can be written as
\ba
\label{X-structure}
X^{ij}({\bf v}) &=& X_L(v) \frac{v^iv^j}{{\bf v}^2} + X_T(v) \Big(\delta^{ij} - \frac{v^iv^j}{{\bf v}^2}\Big) ,
\\[2mm]
\label{Y-structure}
Y^i({\bf v}) &=& Y(v) v^i ,
\ea
where $v \equiv |{\bf v}|$ and the coefficients $ X_L(v), X_T(v)$ and $Y(v)$ are equal to 
\ba
\label{XL-XT-Y}
X_L(v) = \frac{v^iv^j}{{\bf v}^2} X^{ij}({\bf v}) , 
~~~~~~~~~~~
X_T(v) = \frac{1}{2}\Big(\delta^{ij} - \frac{v^iv^j}{{\bf v}^2}\Big) X^{ij}({\bf v}),
~~~~~~~~~~~
Y(v) = \frac{v^i}{{\bf v}^2} Y^i({\bf v}) . 
\ea

The equilibrium distribution function of the form
\be
n^{\rm eq}({\bf p}) \sim \exp \Big(-\frac{E_{\bf p}}{T}\Big) ,
\ee
with $T$ being the temperature of the plasma of light quarks and gluons, where heavy quarks are embedded, is expected to solve the transport equation (\ref{F-K-eq}). This is indeed the case if the coefficients $X^{ij}({\bf v})$ and $Y^i({\bf v})$ obey the condition
\be
\label{eq-X-vs-Y}
X^{ij}({\bf v}) \frac{v^j}{T}  =  Y^i({\bf v}) ,
\ee
which in the isotropic plasma reads
\be
\label{eq-XL-vs-Y}
X_L(v) \frac{1}{T}  =  Y(v).
\ee

When the plasma is isotropic and the coefficients $X_L(v)$ and $X_T(v)$ are equal to each other and independent of $v$, the Fokker-Planck equation reads
\be
\label{F-K-eq-simple}
\bigg( {\cal D} - X \Big(\nabla_p^2 + \frac{1}{T} \nabla_p \cdot {\bf v}  \Big) \bigg) n(t, {\bf r},{\bf p})  = 0,
\ee
where $X \equiv X_L(v) = X_T(v)$.

The quantities $X^{ij}({\bf v})$ and $Y^i({\bf v})$ have a clear physical meaning. As discussed in {\it e.g.} the classical monograph \cite{Kampen:1987}, the average momentum change per unit time and the correlation of momentum changes per unit time are given as
\ba
\label{Dpi-final}
\frac{\langle \Delta p^i \rangle}{\Delta t} 
&=& - Y^i({\bf v}) ,
\\[2mm]
\label{Dpi-Dpj-final}
\frac{\langle \Delta p^i \Delta p^j \rangle}{\Delta t} 
&=& X^{ij}({\bf v}) + X^{ji}({\bf v}) .
\ea

Using the formulas (\ref{Dpi-final}) and (\ref{Dpi-Dpj-final}), $X^{ij}({\bf v})$ and $Y^i({\bf v})$ can be related to the collisional energy loss $\frac{dE}{dx}$ and transverse momentum broadening $\hat{q}$ of a heavy-quark in the quark-gluon plasma, which play an important role in a theoretical description of the jet quenching phenomenon. The parameter $\hat{q}$ controls the radiative energy loss in a plasma medium \cite{Baier:1996sk}. One easily finds that
\ba
\frac{\langle \Delta E \rangle}{\Delta t} 
= \frac{1}{\Delta t} 
\Big\langle \frac{{\bf p} \cdot\Delta {\bf p}}{ \sqrt{m^2 + {\bf p}^2}} \Big\rangle 
= v^i \frac{\langle \Delta p^i \rangle}{\Delta t} .
\ea
Since $\Delta x = v \Delta t $, the energy loss per unit path equals
\be
\label{eloss-Y}
\frac{dE}{dx} =  \frac{v^i}{v} \frac{\langle \Delta p^i \rangle}{\Delta t} = - \frac{v^i}{v} Y^i({\bf v}),
\ee  
which in isotropic plasmas reads 
\be
\label{e-loss-X-Y}
\frac{dE}{dx} = - v Y(v) = -\frac{v}{T} X_L(v) .
\ee 

The coefficient $\hat{q}$, which is the broadening per unit path of the distribution of the test parton's momentum transverse to the initial parton's momentum, is immediately found as
\be
\hat{q} = \frac{1}{v} \Big(\delta^{ij} - \frac{v^i v^j}{{\bf v}^2}\Big) \frac{\langle \Delta p^i \Delta p^j \rangle}{\Delta t} 
=  \frac{1}{v} \Big(\delta^{ij} - \frac{v^i v^j}{{\bf v}^2}\Big) \Big( X^{ij}({\bf v}) +  X^{ji}({\bf v}) \Big) ,
\ee
and in an isotropic plasma it equals
\be
\label{qhat-X-T}
\hat{q} =  \frac{2}{v} \Big(\delta^{ij} - \frac{v^i v^j}{{\bf v}^2}\Big) X^{ij}({\bf v})   = \frac{4}{v} X_T(v) .
\ee

When we deal with an equilibrium plasma and the coefficients $ X_L(v), X_T(v)$ are equal to each other and independent of $v$, the Fokker-Planck equation can be related to the nonrelativistic Langevin equation \cite{Kampen:1987}. Then, the diffusion constant $D$ can be expressed as 
\be
\label{diff-constant-XL-XT}
D = \frac{T^2}{X}.
\ee

We are mostly interested in turbulent QCD plasmas populated with strong chromodynamic fields but we start with the equilibrium system where the fields are only at a level of thermal noise. We rederive the known Fokker-Planck equation \cite{Svetitsky:1987gq} in a different way to demonstrate the reliability of our approach.

\section{Equilibrium plasma}
\label{sec-equilibrium}

We assume that the quark-gluon plasma, in which heavy quarks are embedded, is in thermodynamical equilibrium and we first derive in this section explicit expressions for the coefficients $X_L$, $X_T$ and $Y$ which enter the Fokker-Planck equation.

\subsection{Computation of $X$ and $Y$}
\label{sec-X-Y}

As the formula (\ref{X-def}) shows, the quantity $X$ is given by the correlations functions $\langle E^i(t,{\bf r}) E^j (t',{\bf r}')\rangle$,  $\langle B^i(t,{\bf r}) B^j (t',{\bf r}')\rangle$,  $\langle E^i(t,{\bf r}) B^j (t',{\bf r}')\rangle$,  and $\langle B^i(t,{\bf r}) E^j (t',{\bf r}')\rangle$ which were studied in detail in \cite{Mrowczynski:2008ae}. The explicit  expressions are collected in Appendix \ref{sec-corr-fun}. Since the correlation functions are of the structure (\ref{EE-1}), the tensor $X^{ij}$ is written as
\ba
\label{X-1}
X^{ij}({\bf v}) =  \frac{1}{2N_c} \int_0^t dt'  \int {d\omega \over 2\pi} \int {d^3k \over (2\pi)^3}
e^{-i(\omega - {\bf k} \cdot {\bf v})(t-t')} \langle F^i_a F^j_a \rangle_{\omega,\, {\bf k}},
\ea
where the chromodynamic fields are expressed in the adjoint representation of the ${\rm SU}(N_c)$ group and $\langle F^i_a F^j_a \rangle_{\omega,\, {\bf k}}$ is the fluctuation spectrum. For a translationally invariant system it is defined as
\be
\label{spec-F-def}
\langle F^i_a F^j_a \rangle_{\omega ,\, {\bf k}} 
\equiv \int dt \int d^3r \, e^{i(\omega t - {\bf k}\cdot {\bf r})} 
\langle  F^i_a(t, {\bf r}) \, F^j_a (0, {\bf 0})\rangle .
\ee
A more general definition is discussed in Appendix~\ref{sec-corr-spec}. Combining the equilibrium fluctuation spectra (\ref{EiEj-spec-eq}), (\ref{BiBj-spec-eq}) and (\ref{BiEj-spec-eq}), one finds
\ba
\label{FiFj-spec}
&& \langle F^i_a F^j_a \rangle_{\omega,\, {\bf k}} 
= 2g^2(N_c^2-1)  \frac{\omega^2}{e^{\beta |\omega|} - 1}   \bigg{\{}
 \omega^2 \frac{k^ik^j}{{\bf k}^2}
\frac{\Im \varepsilon_L(\omega,{\bf k})}
{|\omega^2 \varepsilon_L(\omega,{\bf k})|^2}
\\[2mm] \nn
&&+ 
\bigg[\omega \big (v^i k^j + k^i v^j - 2\delta^{ij} ({\bf k}\cdot {\bf v})\big) 
+ {\bf k}^2 
 \Big( \delta^{ij} {\bf v}^2  - v^i  v^j  - \frac{({\bf v} \times {\bf k})^i ({\bf v} \times {\bf k})^j}{{\bf k}^2} \Big) 
+ \omega^2 \Big(\delta^{ij} - \frac{k^ik^j}{{\bf k}^2}\Big) \bigg] 
\frac{\Im \varepsilon_T(\omega,{\bf k})}
{|\omega^2 \varepsilon_T(\omega,{\bf k})-{\bf k}^2|^2} 
 \bigg{\}},
\ea
where $\beta \equiv T^{-1}$ and $\varepsilon_{L,T}(\omega,{\bf k})$ are chromodielectric functions which for the equilibrium plasma of massless particles are also given in Appendix \ref{sec-corr-fun}.

After performing the elementary time integration in Eq.~(\ref{X-1}), one is left with the integral over the four-vector $k^\mu = (\omega, {\bf k})$. Taking into account only the terms of the integrand which are even as a function of $k^\mu$ and give nonzero contributions, one obtains 
\ba
\label{X-3}
X^{ij}({\bf v}) = \frac{1}{2N_c} \int {d\omega \over 2\pi} \int {d^3k \over (2\pi)^3}
\frac{\sin((\omega - \bar\omega)t\big)}{\omega - \bar\omega} 
\langle F^i_a F^j_a \rangle_{\omega,\, {\bf k}} ,
\ea
where $\bar\omega \equiv {\bf k} \cdot {\bf v}$. In the limit $t \rightarrow \infty$, we have
\be
\lim_{t \rightarrow \infty} \frac{\sin \big((\omega - \bar\omega) t\big)}{\omega - \bar\omega} 
= \pi \delta(\omega - \bar\omega),
\ee
and thus we get
\ba
\label{X-FF-final}
X^{ij}({\bf v})
= \frac{1}{4N_c} \int {d^3k \over (2\pi)^3} \langle F^i_a F^j_a \rangle_{\bar\omega ,\, {\bf k}}.
\ea
One can show that the expression (\ref{X-FF-final}) properly approximates the formula (\ref{X-3}) if the spectrum $\langle F^i_a F^j_a \rangle_{\omega,\, {\bf k}}$ weakly changes as a function of $\omega$ in the interval $[\bar\omega - \pi/t, \bar\omega + \pi/t]$. When the time grows the condition is easier and easier to fulfill. 

Since the plasma under consideration is isotropic, the tensor $X^{ij}({\bf v})$ is fully determined by the two functions $X_L(v)$ and $X_T(v)$ introduced in Eq.~(\ref{X-structure}). Substituting the explicit form of the fluctuation spectrum (\ref{FiFj-spec}) into Eq.~(\ref{X-FF-final}), one finds the coefficients $X_L(v)$ and $X_T(v)$ as
\ba
\label{XL-2}
X_L(v) &=&  g^2C_F 
\int {d^3k \over (2\pi)^3}   \frac{|\bar\omega|}{e^{\beta |\bar\omega|} - 1} \frac{\bar\omega ^3}{{\bf v}^2}
\bigg[ 
\frac{\bar\omega^2}{{\bf k}^2}
\frac{\Im \varepsilon_L(\bar\omega,{\bf k})}
{|\bar\omega^2 \varepsilon_L(\bar\omega,{\bf k})|^2}
+ 
\Big({\bf v}^2 - \frac{\bar\omega^2}{{\bf k}^2}\Big) 
\frac{\Im \varepsilon_T(\bar\omega,{\bf k})}
{|\bar\omega^2 \varepsilon_T(\bar\omega,{\bf k})-{\bf k}^2|^2} \bigg]  ,
\\[4mm]
\label{XT-2}
X_T(v) &=&  \frac{g^2C_F}{2}
\int {d^3k \over (2\pi)^3}   \frac{|\bar\omega| \bar\omega }{e^{\beta |\bar\omega|} - 1}\bigg[ 
\bar\omega^2 \Big(1 - \frac{\bar\omega^2}{{\bf v}^2{\bf k}^2}\Big) 
\frac{\Im \varepsilon_L(\bar\omega,{\bf k})}
{|\bar\omega^2 \varepsilon_L(\bar\omega,{\bf k})|^2}
\\[2mm]\nn
&&~~~~~~~~~~~~~~~~~~~~~~~~~~~~~~~~~~~~~~~~~~~
+ \Big({\bf v}^2 {\bf k}^2 - 2 \bar\omega^2
+  \frac{\bar\omega^4}{{\bf v}^2 {\bf k}^2} \Big)
\frac{\Im \varepsilon_T(\bar\omega,{\bf k})}
{|\bar\omega^2 \varepsilon_T(\bar\omega,{\bf k})-{\bf k}^2|^2}  \bigg],
\ea
where $C_F \equiv \frac{N_c^2-1}{2 N_c}$ is the Casimir invariant. 

The coefficient $Y^i({\bf v})$, which is determined by the correlations functions $\langle E^i(t,{\bf r}) \delta Q_0 ({\bf r}', {\bf p}')\rangle$, and $\langle B^i(t,{\bf r}) \delta Q_0 ({\bf r}', {\bf p}')\rangle$ can be derived directly from the formula (\ref{Y-def}). Such a derivation for a simplified case of only longitudinal electric field present in the plasma can be found in \cite{Mrowczynski:2009gf} where it is shown that the condition (\ref{eq-X-vs-Y}) or (\ref{eq-XL-vs-Y}) is indeed satisfied. Here instead we refer to the condition (\ref{eq-X-vs-Y}) to obtain  $Y^i({\bf v})$.

Once the coefficients $X^{ij}({\bf v})$ is given by Eq.~(\ref{X-structure}) together with Eqs.~(\ref{XL-2}), (\ref{XT-2}) and $Y^i({\bf v})$  by Eqs.~(\ref{Y-structure}) and (\ref{eq-XL-vs-Y}), the Fokker-Planck equation (\ref{F-K-eq}) is fully specified. We note that $X^{ij}({\bf v})$ and $Y^i({\bf v})$ depend on heavy quark momentum ${\bf p}$ and its mass $m$ only through the velocity ${\bf v}= {\bf p}/\sqrt{m^2 + {\bf p}^2}$. Therefore, $X_L(v), ~X_T(v)$ and $Y(v)$ become independent of $p$ when the heavy quarks of interest are truly relativistic and ${\bf p}^2 \gg m^2$.  Although, the coefficients $X_L(v), ~X_T(v)$ and $Y(v)$ are independent of the quark mass, the Fokker-Planck equation (\ref{F-K-eq}) does depend  on $m$ which is evident when the momentum derivatives are replaced by the velocity derivatives. 

Since the coefficients $X_L(v), ~X_T(v)$ and $Y(v)$ depend on the quark mass only through the velocity, one might think that the corresponding Fokker-Planck equation is applicable to quarks of any mass. However, it is not true. As mentioned in the introduction, the typical momentum transfer in a single collision, which is of order $gT$, must be much smaller than the quark momentum that is $gT \ll m v/\sqrt{1-v^2}$. And here the quark mass matters. 

\subsection{Limit of small $v$}
\label{sec-FK-eq-small-v}

In the limit of small velocities of heavy quarks, the equilibrium Fokker-Planck equation gets a simpler form and the coefficients $X_L(v), ~X_T(v)$ can be estimated analytically. Indeed, when heavy quarks are nonrelativistic (${\bf v}^2 \ll 1$), we have $\bar{\omega}^2 \ll {\bf k}^2$ and one can use the approximate formulas (\ref{Re-Im-eL-small-omega}) and (\ref{Re-Im-eT-small-omega}). Then, Eqs.~(\ref{XL-2}) and (\ref{XT-2}) read
\ba
\label{XL-small-v-1}
X_L(v) &=&  
\frac{g^2 \pi C_F}{2} \, m_D^2
\int {d^3k \over (2\pi)^3} \frac{\bar\omega^2}{{\bf v}^2 |{\bf k}|}\, 
\frac{1}{(m_D^2 + {\bf k}^2)^2} \frac{|\bar\omega|}{e^{\beta |\bar\omega|} - 1},
\\[4mm]
\label{XT-small-v-1}
X_T(v) &=&  
\frac{g^2\pi C_F}{4} \, m_D^2 
\int {d^3k \over (2\pi)^3} 
\Big(1 - \frac{\bar\omega^2}{{\bf k}^2 {\bf v}^2}\Big) 
\frac{|{\bf k}|}{(m_D^2 + {\bf k}^2)^2} \frac{|\bar\omega|}{e^{\beta |\bar\omega|} - 1},
\ea
where the contributions originating from $\varepsilon_T(\omega,{\bf k})$ appear to vanish. Introducing spherical coordinates with the axis $z$ along the vector ${\bf v}$, the integrals (\ref{XL-small-v-1}) and (\ref{XT-small-v-1}) are rewritten as   
\ba
\label{XL-small-v-2}
X_L(v) &=&  
\frac{g^2 C_F}{4 \pi} \, m_D^2 v
\int_0^\infty \frac{dk \, k^4}{(m_D^2 + k^2)^2}
\int_0^{1}\frac{dx \, x^3}{e^{\beta k x v} - 1} ,
\\[4mm]
\label{XT-small-v-2}
X_T(v) &=&  
\frac{g^2 C_F}{8 \pi} \, m_D^2 v
\int_0^\infty \frac{dk \, k^4}{(m_D^2 + k^2)^2} 
\int_0^{1} \frac{dx (1 - x^2)x}{e^{\beta k x v} - 1} ,
\ea
where the trivial azimuthal integrals are performed and $x \equiv \cos\theta$ with $\theta$ being the angle between ${\bf v}$ and ${\bf k}$. 

When $ T \gg m_D$,  the integrals (\ref{XL-small-v-2}) and (\ref{XT-small-v-2}) can be estimated as follows. One first observes that the dominant contribution comes from $k \in [m_D,T]$. Assuming that $\beta k x v \ll 1$, the integrals over $x$ are easily computed and one obtains
\ba
\label{XL-XT-small-v-3}
X_L(v) =  X_T(v) =  
\frac{g^2 C_F}{12 \pi} \, m_D^2 T \int_{m_D}^T \frac{dk \, k^3}{(m_D^2 + k^2)^2} .
\ea
Approximating the integrand as $k^{-1}$, we finally get 
\ba
\label{XL-XT-small-v-final}
X_L(v) = X_T(v) =  
\frac{g^2 C_F}{12 \pi} \, m_D^2 T  \log\Big(\frac{T}{m_D}\Big)  .
\ea

As one sees in Eq.~(\ref{XL-XT-small-v-3}) or (\ref{XL-XT-small-v-final}), $X_L(v)$ and $X_T(v)$ are independent of $v$ and equal to each other. The formula (\ref{diff-constant-XL-XT}) is therefore applicable and the inverse diffusion constant equals
\be
\label{D-inv-fields}
\frac{1}{D} = \frac{g^2 C_F}{12 \pi} \, \frac{m_D^2}{T} \log\Big(\frac{T}{m_D}\Big),
\ee
which agrees with Eq.~(12) of the study \cite{Moore:2004tg} when only the logarithmic term is taken into account.

\subsection{Numerical results}
\label{sec-numerical-eq}

\begin{figure}[t]
\begin{minipage}{87mm}
\centering
\includegraphics[scale=0.24]{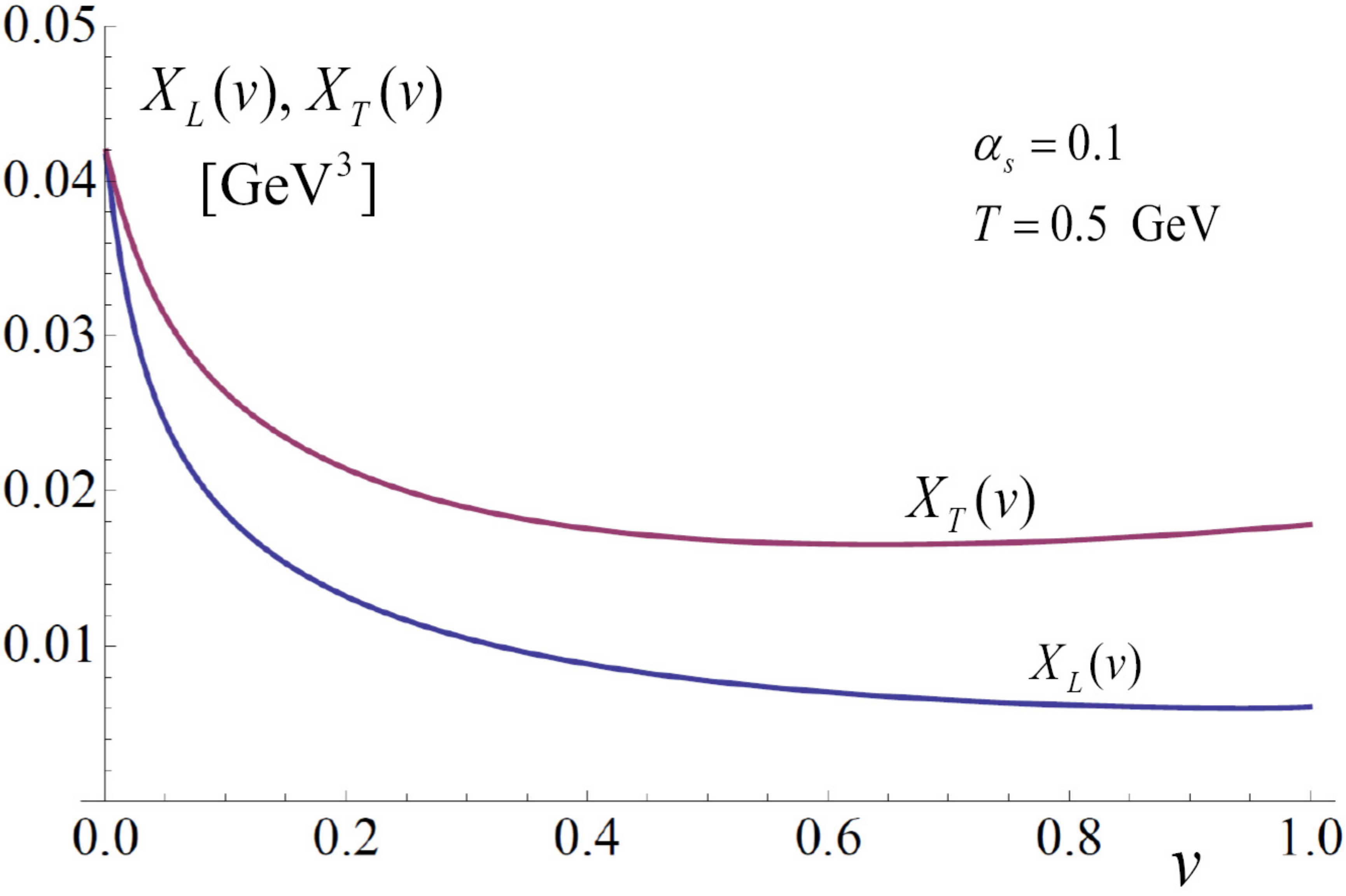}
\caption{The equilibrium coefficients $X_L(v)$ and  $X_T(v)$ as functions of the velocity $v$.}
\label{fig-XL-XT-v-eq}
\end{minipage}
\hspace{1mm}
\begin{minipage}{87mm}
\centering
\includegraphics[scale=0.24]{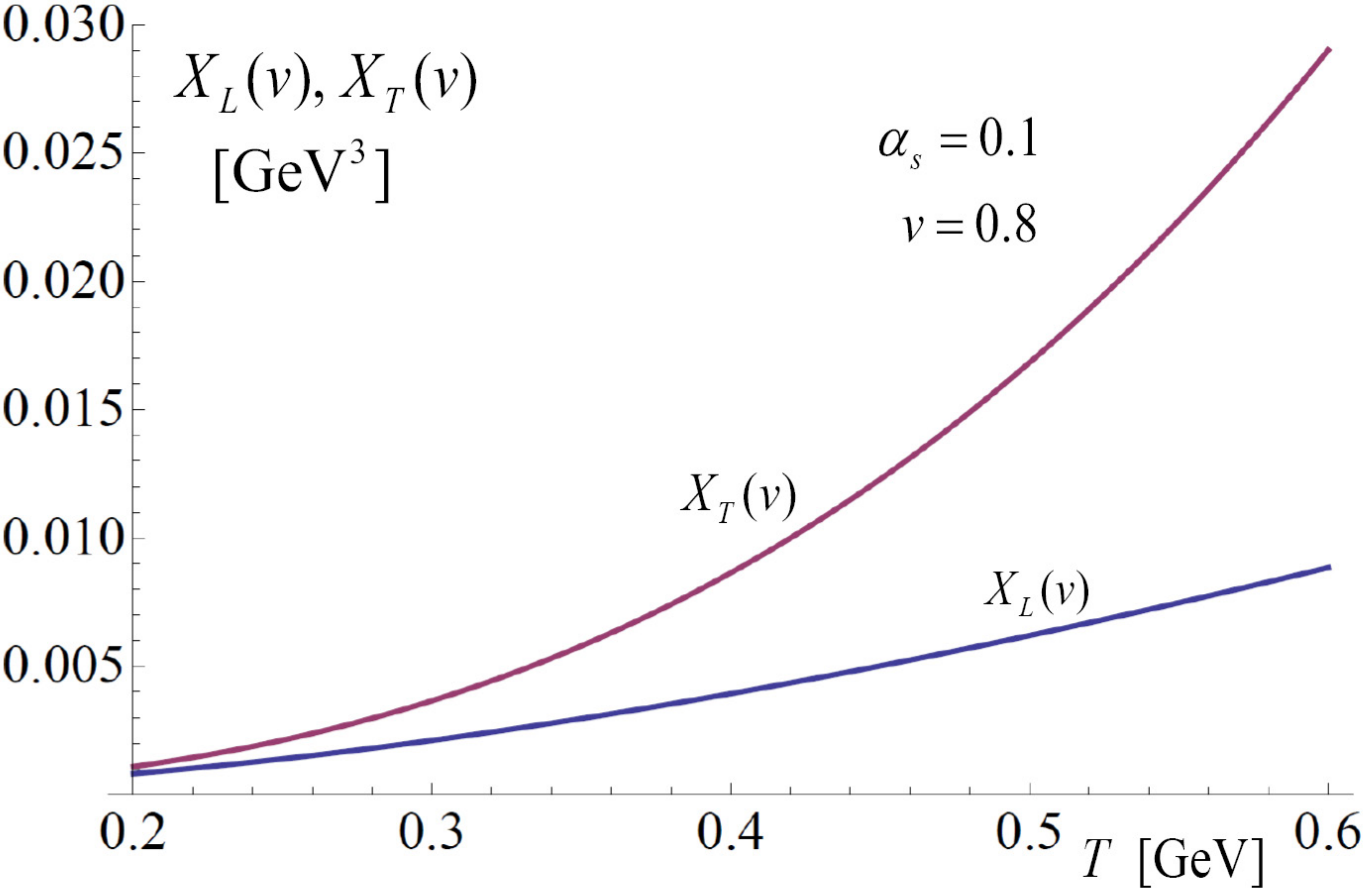}
\caption{The equilibrium coefficients $X_L(v)$ and  $X_T(v)$ as functions of the temperature $T$.}
\label{fig-XL-XT-T-eq}
\end{minipage}
\hspace{2mm}
\end{figure}

Here we show some numerical results for the equilibrium QGP of $N_c=3$ and $N_f =2$. The Debye mass is computed according to the formula (\ref{m_D}). Fig.~\ref{fig-XL-XT-v-eq} presents the coefficients $X_L(v)$ and $X_T(v)$, which are obtained directly from Eqs.~(\ref{XL-2}) and (\ref{XT-2}), as functions of the velocity $v$. The coupling constant and the temperature are $\alpha_s \equiv g^2/4\pi = 0.1$ and  $T = 0.5~{\rm GeV}$. In Fig.~\ref{fig-XL-XT-T-eq} we show how $X_L(v)$ and $X_T(v)$ depend on the temperature $T$. The coupling constant is again $\alpha_s = 0.1$ and the velocity equals $v = 0.8$. 

We have checked that the values of $X_L(v)$ and $X_T(v)$, which we obtained numerically, agree rather well with those computed by Svetitsky \cite{Svetitsky:1987gq} except in the domain of small velocities $v \le 0.1$. The agreement is not trivial because the coefficients of the Fokker-Planck equation were derived in \cite{Svetitsky:1987gq} from the matrix elements of heavy quark binary interactions with plasma constituents. To remove infrared divergences of the matrix elements, a mass parameter, corresponding to the Debye mass, was included in the gluon propagator. Since the procedure is not very accurate, it presumably explains the difference with our $X_L(v)$ and $X_T(v)$ in the domain of small velocities. On the other hand our approach does not treat properly the interactions with a momentum transfer exceeding the Debye mass. Nevertheless the results of both approaches are numerically rather similar.

\section{Turbulent QGP}
\label{sec-turbulent-QGP}

In this section we consider a Fokker-Planck equation of heavy quarks in a turbulent QGP which is populated with strong chromodynamic fields. The plasma is assumed to be isotropic and translationally invariant both in time and space. The tensor $X^{ij}({\bf v})$, which enters the Fokker-Planck equation (\ref{F-K-eq}), is given by Eq.~(\ref{X-FF-final}). The method to obtain $Y^j({\bf v})$, which is used in \cite{Mrowczynski:2009gf}, works only for equilibrium plasmas. Therefore, we will refer to the relation  (\ref{eq-X-vs-Y}). However, it implicitly assumes that in the long time limit the system of heavy quarks described by the Fokker-Planck equation reaches a state of thermal equilibrium with temperature $T$. First of all, the value of $T$ is, in principle, unknown and one needs additional arguments to determine it. There is also a more important problem - it is unclear what are the properties of $X^{ij}({\bf v})$ for which the assumption of equilibrium makes sense. If, for example, there are only magnetic fields in the plasma, $X^{ij}({\bf v})$ is purely transverse and $TY^i({\bf v}) = 0$. Consequently, the Fokker-Planck equation reads
\be
\label{XT-eq}
\frac{\partial}{\partial t} n(t,{\bf p}) = 
\nabla_p^i X_T(v)\Big(\delta^{ij} -\frac{p^i p^j}{{\bf p}^2}\Big) \nabla_p^j  n(t,{\bf p}) ,
\ee
and any stationary isotropic function $n(p)$ solves the equation because $\nabla_p  n(p) \sim {\bf p}$. Therefore, pure magnetic fields do not drive the system to the thermal equilibrium, as expected. In spite of these concerns, we will use the relation  (\ref{eq-X-vs-Y}) to get $Y^j({\bf v})$.

\subsection{Gaussian correlation functions of independent $E$ and $B$ fields}
\label{sec-gauss-EB}

We start with a simple model proposed in \cite{Asakawa:2006jn} where the correlation functions of electric and magnetic fields are chosen in the following Gaussian form
\ba
\label{EE-x-gauss}
\langle E^i_a(t, {\bf r}) \, E^j_b(0, {\bf 0}) \rangle &=& \delta^{ab}\delta^{ij} M_E 
\exp\Big( - \frac{t^2}{2\sigma^2_t} - \frac{{\bf r}^2}{2\sigma^2_r}\Big) ,
\\[2mm]
\label{BB-x-gauss-ind-E-B}
\langle B^i_a(t, {\bf r}) \, B^j_b(0, {\bf 0}) \rangle &=& \delta^{ab}\delta^{ij} M_B 
\exp\Big( - \frac{t^2}{2\sigma^2_t} - \frac{{\bf r}^2}{2\sigma^2_r}\Big) ,
\\[2mm]
\label{EB-BE-x-gauss-ind-E-B}
\langle E^i_a(t, {\bf r}) \, B^j_b(0, {\bf 0}) \rangle &=& \langle B^i_a(t, {\bf r}) \, E^j_b(0, {\bf 0}) \rangle = 0.
\ea
The correlation lengths $\sigma_t, ~\sigma_r$ and the parameters $M_E, ~M_B$ of dimension mass to the fourth power will be discussed later on. The fluctuation spectrum obtained from the correlation function (\ref{EE-x-gauss}) is also Gaussian and it equals
\be
\label{EE-k-gauss}
\langle E^i_a E^j_b \rangle_{\omega ,\, {\bf k}} 
=  \delta^{ab}\delta^{ij} (2\pi)^2 \tau \sigma^3 M_E
\exp\Big( - \frac{\sigma^2_t \omega^2}{2} - \frac{\sigma^2_r {\bf k}^2}{2}\Big) .
\ee

Substituting the correlation functions (\ref{EE-x-gauss}), (\ref{BB-x-gauss-ind-E-B}), and (\ref{EB-BE-x-gauss-ind-E-B}) into Eq.~(\ref{X-FF-final}), the tensor $X^{ij}({\bf v})$ is found as
\ba
\label{X-gauss-E-B-ind}
X^{ij}({\bf v}) =  \sqrt{\frac{\pi}{2}} \, g^2C_F \big(\delta^{ij} M_E + (\delta^{ij} v^2 - v^iv^j) M_B\big)  
\frac{\sigma_t \sigma_r}{\sqrt{\sigma^2_r + v^2 \sigma^2_t}} ,
\ea
which gives
\ba
\label{XL-gauss-E-B-ind}
X_L(v)
&=&   
 \sqrt{\frac{\pi}{2}} \, g^2C_F \,
 \frac{M_E \sigma_t \sigma_r}{\sqrt{\sigma^2_r + v^2 \sigma^2_t}}  ,
\\[2mm]
\label{XT-gauss-E-B-ind}
X_T(v)
&=& 
\sqrt{\frac{\pi}{2}} \, g^2C_F \, \frac{(M_E + v^2 M_B) \sigma_t \sigma_r}{\sqrt{\sigma^2_r + v^2 \sigma^2_t}} .
\ea

When  $v^2 \ll \sigma^2_r/\sigma^2_t$ and $v^2 \ll M_E/M_B$, the coefficients $X_L(v)$ and $X_T(v)$ are, as in the equilibrium case, equal to each other and independent of $v$, and
\be
\label{X-L-X-T-guass-E-B-ind-v-small}
X=X_L(v)=X_T(v) =  \sqrt{\frac{\pi}{2}} \, g^2C_F \, M_E \sigma_t  .
\ee
The Fokker-Planck equation is then of the form (\ref{F-K-eq-simple}).

\subsection{Gaussian correlation function of vector potentials}
\label{sec-gauss-A}

Since the electric and magnetic fields are, in general, coupled to each other, the functions $\langle E_a^i \, E^j_b \rangle$, $\langle E_a^i \, B^j_b \rangle$, $\langle B_a^i \, E^j_b \rangle$, and $\langle B_a^i \, B^j_b \rangle$ are not fully independent from each other. The electric and magnetic fields are automatically related to each other if one postulates the correlation function of the four-potential and  then computes the correlation functions of the $E-$ and $B-$fields. In this section we follow this path. Specifically, we assume the Gaussian correlation function of the vector potential 
\be
\label{AA-x-gauss}
\langle A^i_a(t, {\bf r}) \, A^j_b(0, {\bf 0}) \rangle = \delta^{ab}\delta^{ij} M_A 
\exp\Big( - \frac{t^2}{2\sigma^2_t} - \frac{{\bf r}^2}{2\sigma^2_r}\Big) .
\ee
The parameter $M_A$ of the dimension mass squared will be discussed later on. The fluctuation spectrum of the potential equals
\ba
\label{AA-k-gauss}
\langle A^i_a A^j_b \rangle_{\omega ,\, {\bf k}} 
=  \delta^{ab}\delta^{ij} (2\pi)^2 \sigma_t \sigma^3_r M_A
\exp\Big( - \frac{\sigma^2_t\omega^2}{2} - \frac{\sigma^2_r{\bf k}^2}{2}\Big) .
\ea
We further choose the radiation gauge
\be
\label{radiation-gauge}
A^0_a(t, {\bf r}) = 0, ~~~~~~~~~~~
\nabla \cdot {\bf A}_a(t, {\bf r}) = 0 ,
\ee
and the electric and magnetic fields are obtained in {\em the linear regime} as
\be
\label{E-B-vs-A-linear}
{\bf E}_a(t, {\bf r}) = - \dot{\bf A}_a(t, {\bf r}), ~~~~~~~~~~~
{\bf B}_a(t, {\bf r}) = \nabla \times {\bf A}_a(t, {\bf r}) .
\ee

To get the fluctuation spectra $\langle E^i_a E^j_b\rangle_{\omega, {\bf k}}$, $\langle B^i_a E^j_b\rangle_{\omega, {\bf k}}$, $\langle B^i_a B^j_b\rangle_{\omega, {\bf k}}$ from the spectrum (\ref{AA-k-gauss}) we refer to the relation (\ref{fluc-spec-vs-FT}) derived in Appendix~\ref{sec-corr-spec}. Using Eq.~(\ref{E-B-vs-A-linear}), the fluctuation spectra are found as
\ba
\label{EE-k-AA-gauss}
\langle E^i_a E^j_b\rangle_{\omega, {\bf k}} 
&=&  \delta^{ab}  (2\pi)^2 \sigma_t \sigma^3_r M_A \, \omega^ 2 \delta^{ij}
\exp\Big( - \frac{\sigma^2_t\omega^2}{2} - \frac{\sigma^2_r{\bf k}^2}{2}\Big)  ,
\\[2mm]
\label{BB-k-AA-gauss}
\langle B^i_a B^j_b \rangle_{\omega, {\bf k}} 
&=& 
\delta^{ab}  (2\pi)^2 \sigma_t \sigma^3_r M_A
\Big(\delta^{ij} {\bf k}^2 - k^i k^j \Big)
\exp\Big( - \frac{\sigma^2_t\omega^2}{2} - \frac{\sigma^2_r{\bf k}^2}{2}\Big) ,
\\[2mm]
\label{EB-k-AA-gauss}
\langle E^i_a B^j_b\rangle_{\omega, {\bf k}} 
=  \langle B^j_a E^i_b\rangle_{\omega, {\bf k}} &=&
\delta^{ab}  (2\pi)^2 \sigma_t \sigma^3_r M_A \, \omega \,
\epsilon^{jmi} k^m 
\exp\Big( - \frac{\sigma^2_t\omega^2}{2} - \frac{\sigma^2_r{\bf k}^2}{2}\Big) .
\ea

Substituting the formulas (\ref{EE-k-AA-gauss}), (\ref{BB-k-AA-gauss}), and (\ref{EB-k-AA-gauss}) into Eq.~(\ref{X-FF-final}), one gets after some manipulations
\ba
\label{X-mod-Gauss}
X^{ij}({\bf v})
&=&  2\pi^2 g^2C_F  \sigma_t \sigma^3_r M_A 
\int \frac{d^3k}{(2\pi)^3}  \,  
\exp\Big(- \frac{\sigma^2_t\omega^2}{2} - \frac{\sigma^2_r{\bf k}^2}{2}\Big) 
\\[2mm] \nn
&&~~~~~~~~~~~~~~~~~~
\times \Big[(k^i v^j + k^j v^i)\bar\omega - \delta^{ij} \bar\omega^2
+ (\delta^{ij} {\bf v}^2 - v^i v^j){\bf k}^2 - ({\bf k} \times {\bf v})^i ({\bf k} \times {\bf v})^j  \Big] ,
\ea
which gives
\ba
\label{X-L-Guass}
X_L(v) &=& \sqrt{\frac{\pi}{2}} \, g^2C_F \, 
\frac{M_A \sigma_t \sigma_r v^2}{(\sigma^2_r + v^2 \sigma^2_t)^{3/2}} ,
\\[4mm]
\label{X-T-Guass}
X_T(v) &=& \sqrt{\frac{\pi}{8}} \, g^2C_F \, 
\frac{M_A \,\sigma_t v^2}{\sigma_r}
\bigg[  \frac{3}{(\sigma^2 + v^2 \tau^2)^{1/2}} 
- \frac{\sigma^2 + 3 \tau^2 v^2}{(\sigma^2 + v^2 \tau^2)^{3/2}} \bigg] .
\ea

When $v^2 \ll \sigma^2_r/\sigma^2_t$, we find again, as in the equilibrium case,  that the coefficients $X_L(v)$, $X_T(v)$ are equal to each other, but they do depend on $v$. Specifically,  
\be
\label{X-L-X-T-mod-Guass-v-small}
X_L(v) = X_T(v) 
= \sqrt{\frac{\pi}{2}} \, g^2C_F \, \frac{M_A \sigma_t v^2}{\sigma^2_r} .
\ee
In contrast to the equilibrium result and Eq.~(\ref{X-L-X-T-guass-E-B-ind-v-small}), $X_L(v)$ and $X_T(v)$ vanish as $v \to 0$.

\subsection{Stationary power spectrum of vector potential}
\label{sec-stationary-A}

When the momentum distribution of plasma constituents is anisotropic, the system is unstable due to the Weibel instability, see the review \cite{Mrowczynski:2016etf}, and a strong chromomagnetic field is generated spontaneously.  As shown in the numerical study \cite{Arnold:2005ef}, the fluctuation spectrum of the soft fields becomes up to the wave number $k_{\rm max}$ stationary after a sufficiently long time and the spectrum decays with wave vector as $k^{-2}$.  Inspired by these findings we choose, as previously, the radiation gauge (\ref{radiation-gauge}) and consider the following spectrum
\ba
\label{AA-k-power}
\langle A^i_a A^j_b \rangle_{\omega ,\, {\bf k}} 
=  \delta^{ab}\delta^{ij} 
\frac{ 2\pi \delta (\omega) }{\mu^2 + {\bf k}^2} \, \Theta(|{\bf k}| - k_{\rm max})\, {\cal M} ,
\ea
where the parameters ${\cal M}$, $\mu$ and $k_{\rm max}$, which are all of the dimension of mass, will be determined later on. We note that $k_{\rm max}$ is of order of the Debye mass and that the small but nonzero parameter $\mu$ is introduced to eliminate the infrared divergence of the expression (\ref{AA-k-power}). In contrast to the study \cite{Arnold:2005ef}, there are only zero frequency modes in  the spectrum (\ref{AA-k-power}), and consequently the correlators involving electric field vanish. The only non-vanishing correlator is  
\ba
\label{BB-k-power}
\langle B^i_a B^j_b \rangle_{\omega ,\, {\bf k}} 
=  \delta^{ab}\Big(\delta^{ij} {\bf k}^2 - k^i k^j \Big)
\frac{ 2\pi \delta (\omega) }{\mu^2 + {\bf k}^2} \, \Theta(|{\bf k}| - k_{\rm max})\, {\cal M} .
\ea

Using the formula (\ref{X-1}), the tensor $X^{ij}({\bf v})$ is found as
\ba
\label{X-power}
X^{ij}({\bf v}) =
\frac{g^2C_F}{2} \int {d^3k \over (2\pi)^3} 
\Big( \big(\delta^{ij}{\bf v}^2 -v^iv^j\big) {\bf k}^2 - ({\bf v} \times {\bf k})^i ({\bf v} \times {\bf k})^j \Big)
\frac{ 2\pi \delta (\bar\omega) }{\mu^2 + {\bf k}^2} \, \Theta(|{\bf k}| - k_{\rm max})\, {\cal M}.
\ea
The coefficient $X_L(v)$ given by Eq.~(\ref{XL-XT-Y}) vanishes because of transversality of magnetic field while $X_T(v)$ equals
\ba
\label{X-T-power-0}
X_T(v) =
\frac{g^2C_F}{4} {\cal M}\,{\bf v}^2 \int {d^3k \over (2\pi)^3} 
 {\bf k}^2 \frac{ 2\pi \delta (\bar\omega) }{\mu^2 + {\bf k}^2} \, \Theta(|{\bf k}| - k_{\rm max}) .
\ea
After taking the elementary integrals, one obtains
\ba
\label{X-T-power}
X_T(v) = 
\frac{g^2C_F}{8\pi} {\cal M}v\, 
\bigg[\frac{1}{2} k_{\rm max}^2 - \frac{1}{2} \mu^2 \ln\Big(\frac{ k_{\rm max}^2 +\mu^2}{\mu^2} \Big)\bigg] .
\ea
As seen, $X_T(v)$ vanishes when $v \to 0$ and the formula (\ref{X-T-power}) simplifies to  
\ba
\label{X-T-power-1}
X_T(v) = \frac{g^2C_F}{16 \pi} {\cal M}\, k_{\rm max}^2  v ,
\ea
when $k_{\rm max} \gg \mu$.

In Tables~\ref{table-XL-XT} and \ref{table-trans-coef} there are collected the coefficients $X_L(v)$, $X_T(v)$ and corresponding transport coefficients $\frac{dE}{dx}$, $\hat{q}$ obtained in the models of turbulent plasma discussed in the subsections \ref{sec-gauss-EB}, \ref{sec-gauss-A} and \ref{sec-stationary-A}. The models are called `Gauss E \& B', `Gauss A' and `Stationary A', respectively.  We observe that $\frac{dE}{dx}$ is not always reliable because there is no guarantee that heavy quarks evolve in turbulent plasmas towards the state of thermodynamical equilibrium as required by the  relation (\ref{eq-XL-vs-Y}) that is used. We also note that the formula of $\hat{q}$ in the `Gauss E \& B' model, which holds not for a test quark with any $v$ but for a test gluon with $v=1$, was first obtained in  \cite{Majumder:2007zh}.

\begin{table}[b]
\caption{\label{table-XL-XT} The coefficients $X_L(v)$ and $X_T(v)$ in the models of turbulent plasmas}
\center
\begin{tabular}{c  c  c  c  c c c}
\hline \hline 
\noalign{\smallskip}
&~~~~~~& $\frac{1}{g^2 C_F} X_L(v)$ &~~~~~~& $ \frac{1}{g^2 C_F} X_T(v)$  &~~~~~~ &
\\[2mm]
\hline
\noalign{\smallskip}
Gauss E \& B & & $\sqrt{\frac{\pi}{2}}  \frac{M_E \sigma_t \sigma_r}{\sqrt{\sigma_r^2 + v^2 \sigma^2_t}}$ 
&&  $\sqrt{\frac{\pi}{2}}  \frac{(M_E + v^2 M_B) \sigma_t \sigma_r}{\sqrt{\sigma^2_r + v^2 \sigma^2_t}}$
\\[2mm]
\hline
\noalign{\smallskip}
Gauss A & & $ \sqrt{\frac{\pi}{2}} \frac{M_A \sigma_t \sigma_r v^2}{(\sigma^2_r + v^2 \sigma^2_t)^{3/2}}$ 
&&  $\sqrt{\frac{\pi}{8}} \, \frac{M_A \,\sigma_t v^2}{\sigma_r} \Big[ \frac{3}{(\sigma^2_r + v^2 \sigma^2_t)^{1/2}} 
- \frac{\sigma^2_r + 3 \sigma^2_t v^2}{(\sigma^2_r + v^2 \sigma^2_t)^{3/2}} \Big] $
\\[2mm]
\hline
\noalign{\smallskip}
Stationary A & &  0  &&  $\frac{ {\cal M} v}{8\pi} \big[\frac{1}{2} k_{\rm max}^2 
- \frac{1}{2} \mu^2 \ln\big(\frac{ k_{\rm max}^2 +\mu^2}{\mu^2} \big)\big] $
\\[2mm]
\hline
\end{tabular}
\end{table}

\section{Equilibrium vs. isotropic turbulent plasma}
\label{sec-eq-vs-turb}

\begin{figure}[t]
\begin{minipage}{87mm}
\centering
\includegraphics[scale=0.24]{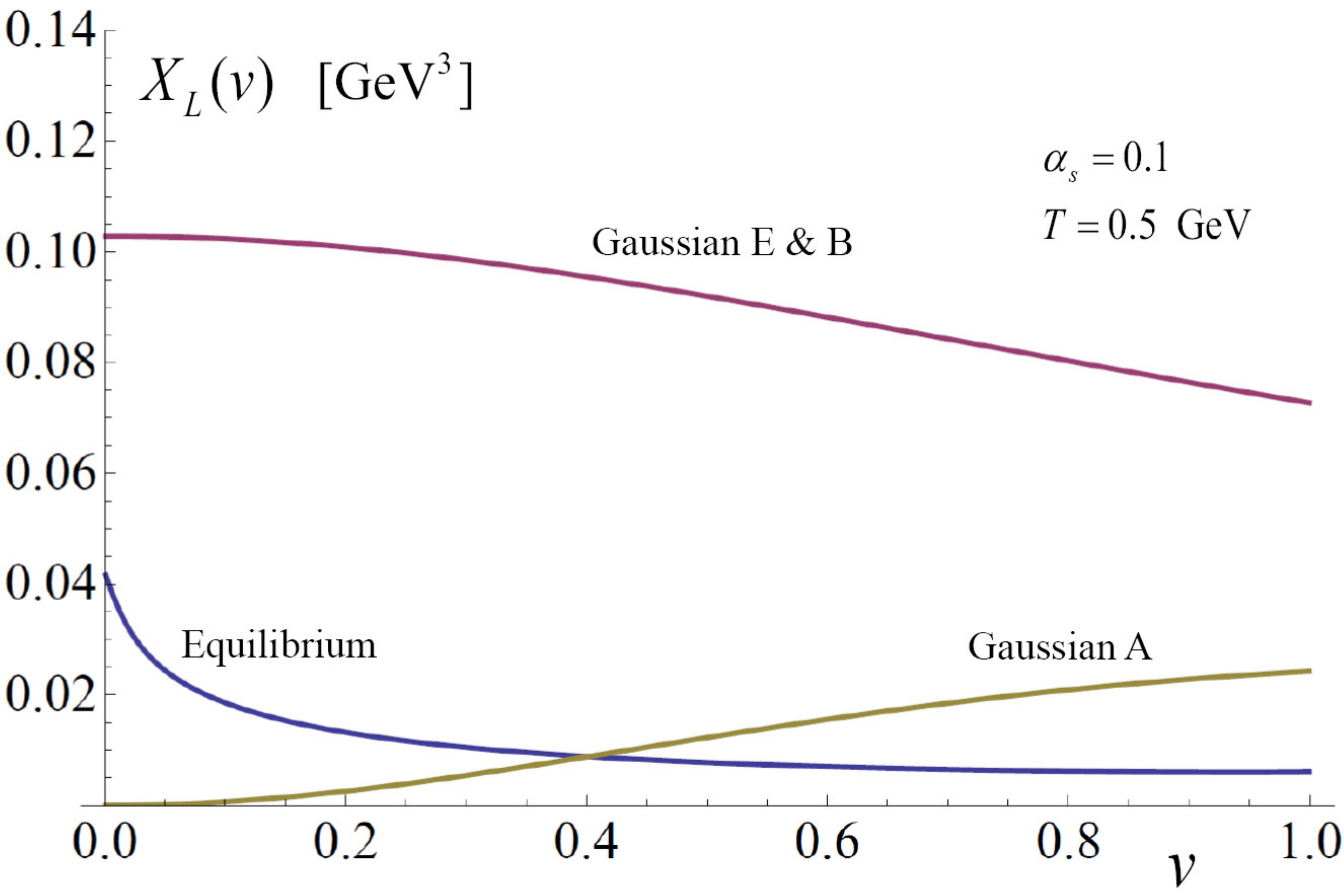}
\caption{The coefficient $X_L(v)$ as a function of the velocity $v$ in the equilibrium plasma and the models of turbulent plasma.}
\label{fig-XL}
\end{minipage}
\hspace{1mm}
\begin{minipage}{87mm}
\centering
\includegraphics[scale=0.235]{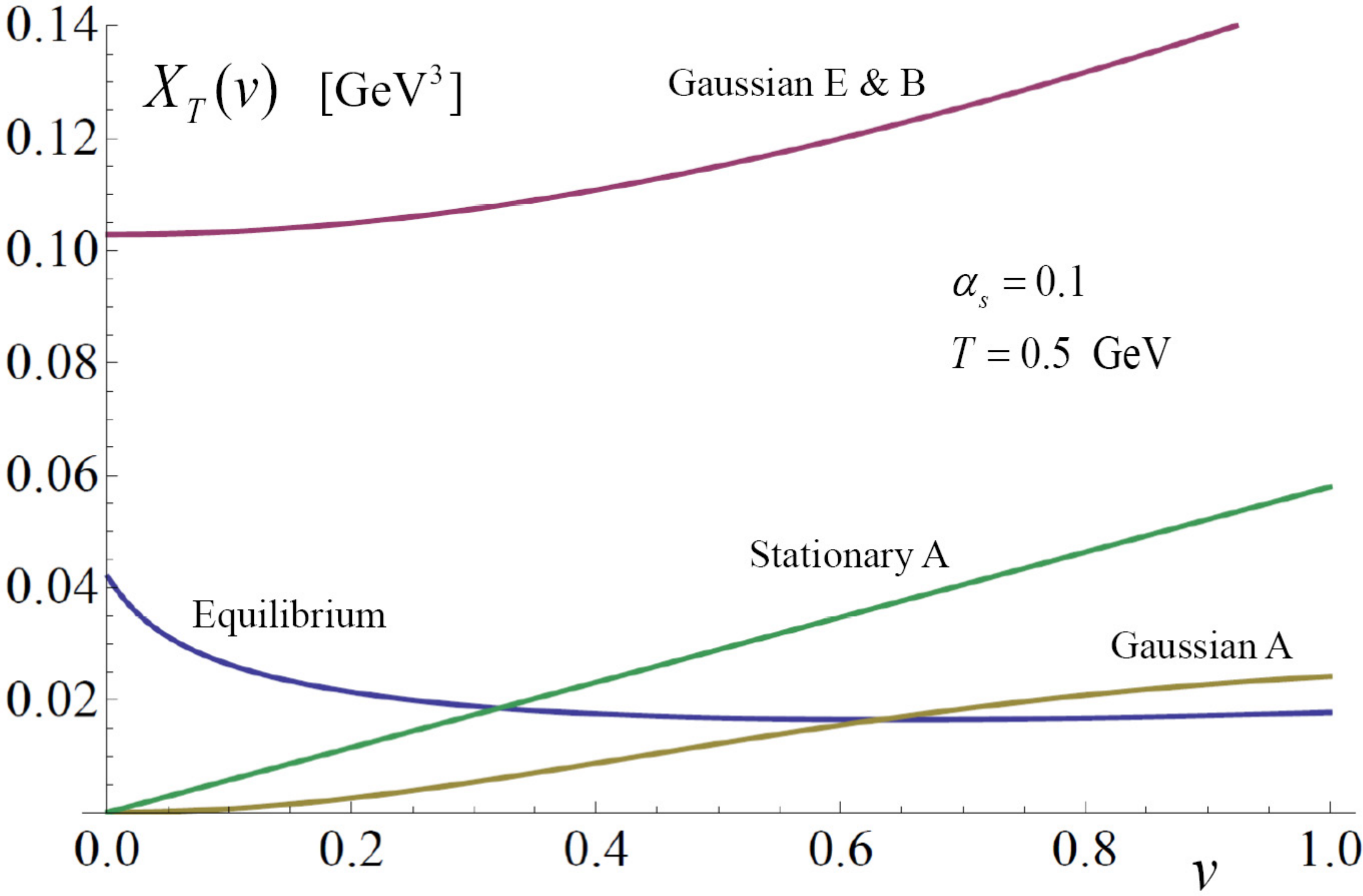}
\caption{The coefficient $X_T(v)$ as a function of the velocity $v$ in the equilibrium plasma and the models of turbulent plasma.}
\label{fig-XT}
\end{minipage}
\hspace{2mm}
\end{figure}

\begin{table}[b]
\caption{\label{table-trans-coef} The transport coefficients  $\frac{dE}{dx}$ and $\hat{q}$ in the models of turbulent plasmas.}
\center
\begin{tabular}{c  c  c  c  c c c}
\hline \hline 
\noalign{\smallskip}
&~~~~~~& $-\frac{1}{g^2 C_F}\, \frac{dE}{dx} $ &~~~~~~& $\frac{1}{g^2 C_F}\, \hat{q} $  
\\[2mm]
\hline \hline 
\noalign{\smallskip}
Gauss E \& B & & $ \sqrt{\frac{\pi}{2}}\frac{M_E \sigma_t \sigma_r v}{(\sigma^2_r + v^2 \sigma^2_t)^{1/2}T} $ 
&&  $\sqrt{8\pi}\, \frac{(M_E + v^2 M_B)\sigma_t \sigma_r}{v\sqrt{\sigma^2_r + v^2 \sigma^2_t}}$
\\[2mm]
\hline
\noalign{\smallskip}
Gauss A & &$ \sqrt{\frac{\pi}{2}}\frac{M_A \sigma_t \sigma_r v^3}{(\sigma^2_r + v^2 \sigma_t^2)^{3/2}T}$ 
&& $\sqrt{\frac{\pi}{2}} \frac{M_A \sigma_t v}{\sigma_r}
\Big[ \frac{3}{(\sigma^2_r + v^2 \sigma^2_t)^{1/2}} 
- \frac{\sigma^2_r + 3 v^2 \sigma^2_t}{(\sigma^2_r + v^2 \sigma^2_t)^{3/2}} \Big] $
\\[2mm]
\hline
\noalign{\smallskip}
Stationary A & & 0 &&  $\frac{ {\cal M}}{2\pi} \big[\frac{1}{2} k_{\rm max}^2 
- \frac{1}{2} m^2 \ln\big(\frac{ k_{\rm max}^2 +m^2}{m^2} \big)\big] $
\\[2mm]
\hline
\end{tabular}
\end{table}

We are going to compare the equilibrium plasma to the turbulent one at the same energy density. The energy density of the equilibrium quark-gluon plasma of $N_f$ massless flavors equals
\be
\label{en-den-QGP}
\varepsilon_{\rm QGP} = \frac{\pi^2}{60} \big( 4(N_c^2 -1) + 7 N_f N_c \big)  T^4 .
\ee

The density of energy accumulated in chromodynamic fields is expressed as 
\ba
\label{en-den-field}
\varepsilon_{\rm field} = \frac{1}{2} \langle E^i_a(t, {\bf r}) \, E^i_a(t, {\bf r}) \rangle 
+ \frac{1}{2} \langle B^i_a(t, {\bf r}) \, B^i_a(t, {\bf r}) \rangle 
= \frac{1}{2} \int \frac{d \omega}{2\pi} \frac{d^3 k}{(2\pi)^3}
\Big(\langle E^i_a E^i_a \rangle_{\omega ,\, {\bf k}} 
+ \langle B^i_a B^i_a \rangle_{\omega ,\, {\bf k}} \Big).
\ea
If the fluctuation spectra $\langle E^i_a E^i_a \rangle_{\omega ,\, {\bf k}}$ and $\langle B^i_a B^i_a \rangle_{\omega ,\, {\bf k}}$ are of the Gaussian form (\ref{EE-k-gauss}), the energy density equals
\ba
\label{en-den-E-B-Gauss}
\varepsilon_{\rm field} = \frac{3(N_c^2-1)}{2} (M_E + M_B) .
\ea
When the fluctuation spectra $\langle E^i_a E^i_a \rangle_{\omega ,\, {\bf k}}$ and $\langle B^i_a B^i_a \rangle_{\omega ,\, {\bf k}}$ are given by Eqs.~(\ref{EE-k-AA-gauss}) and (\ref{BB-k-AA-gauss}), we have
\ba
\label{en-den-A-Gauss}
\varepsilon_{\rm field} =  \frac{3(N_c^2-1)}{2} \Big(\frac{1}{\sigma^2_t} + \frac{2}{\sigma^2_r}\Big) M_A .
\ea
Finally, if the fluctuation spectrum $\langle A^i_a A^i_a \rangle_{\omega ,\, {\bf k}}$ is of the form (\ref{AA-k-power}), the purely magnetic energy density equals
\ba
\label{en-den-A-power}
\varepsilon_{\rm field}  = \frac{N_c^2-1 }{6\pi^2}  \, {\cal M} \, k_{\rm max}^3 = \frac{4}{3\pi^2}  \, {\cal M} \, k_{\rm max}^3,
\ea
where we assume that $k_{\rm max} \gg \mu$. We note that in the weak coupling limit the magnitudes of electric and magnetic fields in turbulent plasmas, which are according to the above estimates of the order $T^2$, are much larger than in equilibrium plasmas where the fields are typically of the order $g^2T^2$. This is the main reason why the turbulent plasmas, which are discussed here, are qualitatively different than the equilibrium one. 

\begin{figure}[t]
\begin{minipage}{87mm}
\centering
\includegraphics[scale=0.24]{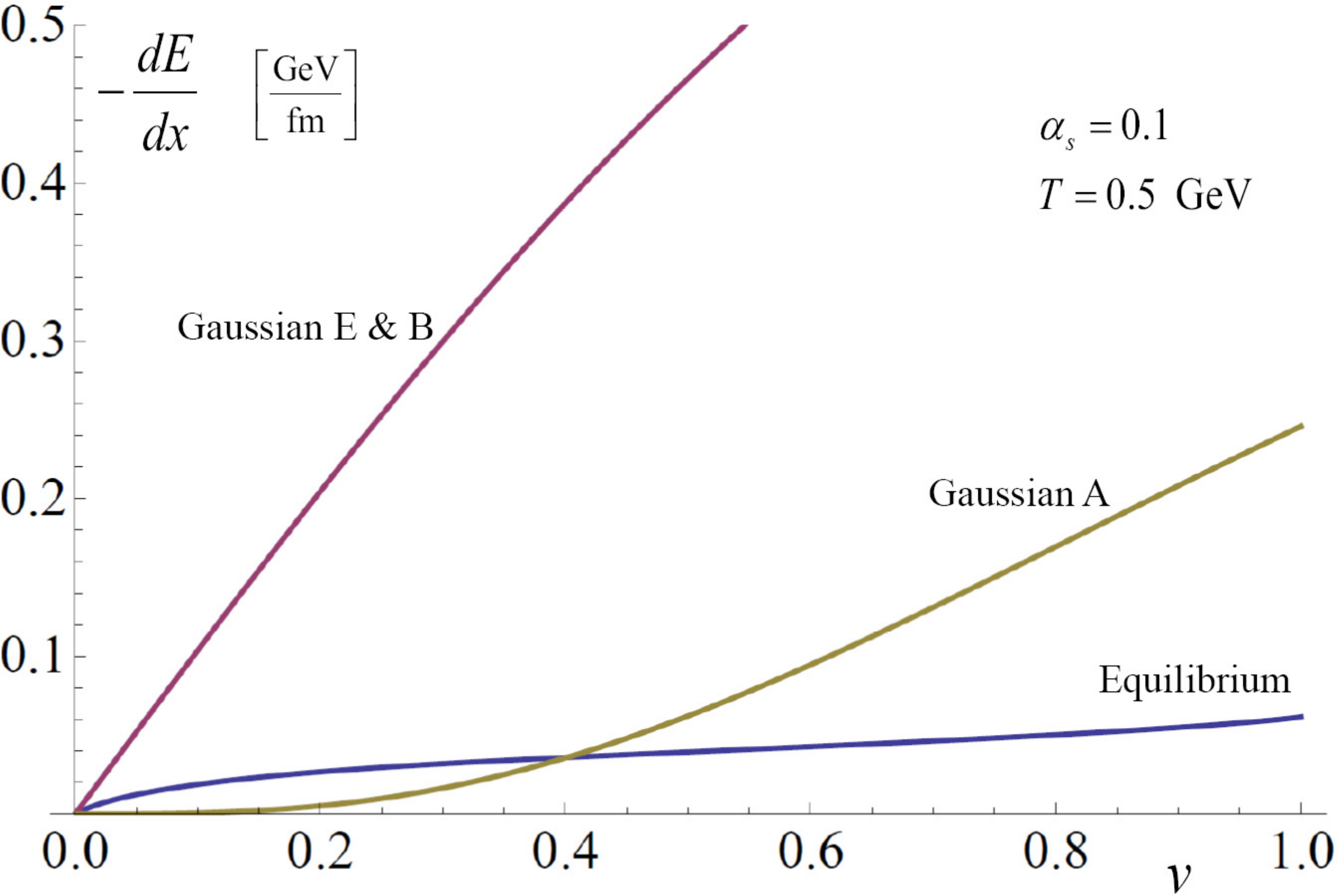}
\caption{The energy loss as a function of $v$ in the models of turbulent plasma.}
\label{fig-eloss}\end{minipage}
\hspace{1mm}
\begin{minipage}{87mm}
\centering
\includegraphics[scale=0.24]{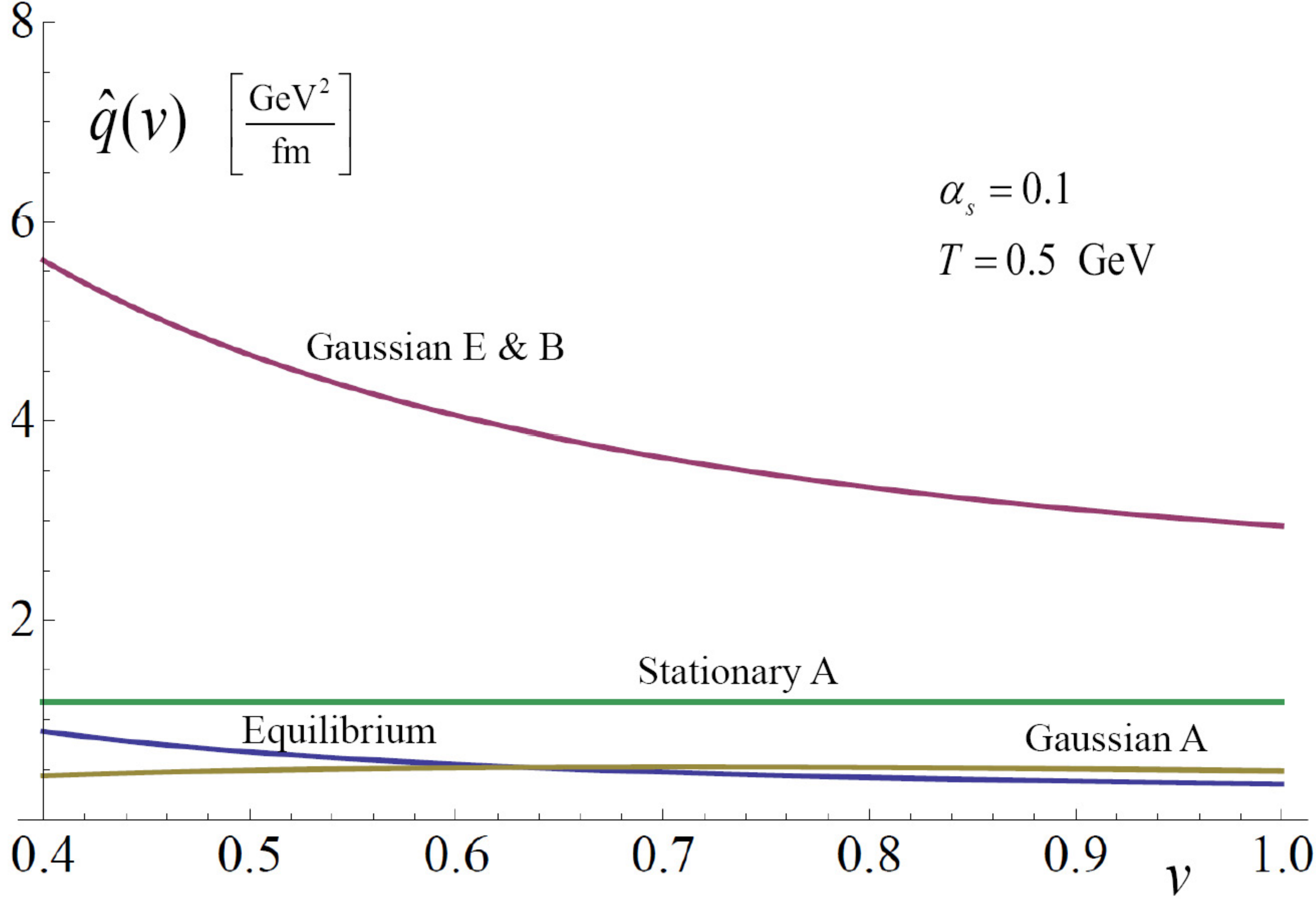}
\caption{The momentum broadening $\hat{q}$ in the models of turbulent plasma.}
\label{fig-qhat}
\end{minipage}
\end{figure}

To reduce the number of parameters of our models of turbulent plasma we adopt the simplifying assumptions that $M_E = M_B$ and $\sigma_t = \sigma_r$. Then, when the energy density of the fields in turbulent plasma (\ref{en-den-E-B-Gauss}), (\ref{en-den-A-Gauss}) and (\ref{en-den-A-power}) equals the energy density of the equilibrium QGP (\ref{en-den-QGP}), the parameters $M_E$, $M_A$ and ${\cal M}$ are equal to
\be
\label{M-esti}
M_E = \frac{37\pi^2}{720}   T^4 ,
~~~~~~~~~~~~~~~~
M_A = \frac{37\pi^2}{1080} \sigma^2_t  T^4 ,
~~~~~~~~~~~~~~~~ 
{\cal M} = \frac{37\pi^4}{40}  \frac{T^4}{k_{\rm max}^3} .
\ee
We set $N_c=3$ and $N_f=2$. With the simplifying relations $M_E = M_B$ and $\sigma_t = \sigma_r$ and the equalities (\ref{M-esti}), one obtains the parametric estimates of the coefficient $X_T(v)$ which are shown in the second column of Table~\ref{table-XT-parametric} where we have additionally assumed that $v^2 \ll 1$. 

Since the parameters $\sigma_t,~\sigma_r$ determine the field correlation length in time and space, they are of order of the screening length and thus we choose $\sigma_t = \sigma_r = m_D^{-1}$. Similarly, $k_{\rm max}$ is a soft momentum and following \cite{Arnold:2005ef} we set $k_{\rm max} = 5 m_D$.  Using the formula (\ref{m_D}), we get 
\be
\label{tau-kmax-esti}
\sigma_t = \frac{\sqrt{3}}{2 g T} , ~~~~~~~~~~~~~~~~~~
k_{\rm max} = \frac{10}{\sqrt{3}}\, g T .
\ee
The parametric estimates of $X_T(v)$, which take into account the relations (\ref{tau-kmax-esti}), are shown in the third column of Table~\ref{table-XT-parametric}. As seen, the interaction of heavy quarks is much stronger in the turbulent plasma than in the equilibrium one if the plasma is truly weakly coupled with $g \ll 1$. The effect disappears when the coupling constant $\alpha_s \equiv g^2/4\pi$ is of realistic value and gives $g$ close to unity or even bigger. 

One asks whether the strong magnetic fields discussed above are stable or may be they rapidly decay due to the Nielsen-Olesen instability \cite{Nielsen:1978rm} which has been discussed in the context of quark-gluon plasma in \cite{Berges:2011sb,Kurkela:2012hp}. Strictly speaking, the Nielsen-Olsen instability occurs in systems with a homogeneous magnetic field because the system's energy can be reduced by particles circulating in the magnetic fields. The unstable mode grows as $e^{\gamma t}$ with $\gamma = \sqrt{gB}$. When the field is inhomogeneous with the wave vector $k$, the decrement of growth is reduced to $\gamma = \sqrt{gB - k^2}$. The instability disappears when $gB < k^2$ which physically means that the length of inhomogeneity is shorter than the Larmor radius. As discussed above, $B \sim T^2$ and $k \sim gT$ in the turbulent plasma and thus the Nielsen-Olsen instability is unavoidable as long as $g \ll 1$. The lifetime of strong magnetic is then of order $(\sqrt{g}\,T)^{-1}$. In Sec.~\ref{sec-glasma} we discuss the glasma \cite{Gelis:2012ri} which is of our main interest. We show that the instability is absent in the glasma because the field correlation length is of the order of the inverse stauration scale $Q_s$. 

In Figs.~\ref{fig-XL} and \ref{fig-XT} we present how $X_L(v)$ and $X_T(v)$ depend on the velocity $v$ in the plasma models under consideration. The coupling constant and temperature are $\alpha_s =0.1$ and $T = 0.5 ~{\rm GeV}$. Figs.~\ref{fig-eloss} and \ref{fig-qhat} show the corresponding energy loss and momentum broadening as functions of $v$. Since the energy density of the turbulent plasma is fixed, the temperature $T$, which enters Eq.~ (\ref{eq-X-vs-Y}) or (\ref{eq-XL-vs-Y}), is fixed as well. The coefficient $\hat{q}$ is computed for $v \ge 0.3$ because it diverges as $v \to 0$ in the model `Gauss E \& B'. However, $\hat{q}$ is of no physical relevance for small velocities. As seen, the coefficients  $X_L(v)$ and $X_T(v)$ and consequently $dE/dx$ and $\hat{q}$ are rather different in the equilibrium and turbulent plasmas - not only the magnitudes differ but the dependences on the heavy quark's velocity are different. The interaction of heavy quarks is particularly strong in the model `Gauss E \& B'. 

One wonders why the models `Gauss E \& B' and `Gauss A', which at first glance are quite similar, give rather different results shown Figs.~\ref{fig-XL} and \ref{fig-XT}. Comparing the fluctuation spectra (\ref{EE-k-AA-gauss}) and (\ref{BB-k-AA-gauss}) to (\ref{EE-k-gauss}), one observes that the low frequency and long wavelength fields are suppressed in the `Gauss A' model. These fields appear to contribute more effectively to the transport coefficients than the high frequency and short wavelength fields.

\begin{table}[b]
\caption{\label{table-XT-parametric} Parametric estimates of the coefficients $X_T(v)$ in the four plasma models under consideration. The third column takes into account that $\sigma_t^{-1} \sim k_{\rm max} \sim g T$.}
\center
\begin{tabular}{c  c  c  c  c c c}
\hline \hline 
\noalign{\smallskip}
&~~~~~~& $X_T(v)$ &~~~~~~& $X_T(v)$  &~~~~~~ &
\\[2mm]
\hline \hline 
\noalign{\smallskip}
equilibrium & & $g^4 T^3$     & & $g^4 T^3$
\\[2mm]
\hline
\noalign{\smallskip}
Gauss E \& B & & $g^2 \tau T^4$ &&  $gT^3$
\\[2mm]
\hline
\noalign{\smallskip}
Gauss A & & $g^2 \tau T^4v^2$ && $gT^3 v^2$
\\[2mm]
\hline
\noalign{\smallskip}
Stationary A & & $g^2\frac{T^4}{k_{\rm max}} v$ && $gT^3 v$
\\[2mm]
\hline
\end{tabular}
\end{table}

\section{Glasma}
\label{sec-glasma}

When relativistic heavy ions collide color charges of partons confined in the colliding nuclei generate strong chromodynamic fields right after the collision. Since the system of infinitely contracted nuclei moving against each other with the speed of light is boost invariant, so is the configuration of generated fields. As shown in the detailed analytic study \cite{Chen:2015wia}, the chromoelectric and chromomagnetic fields spanned between the receding nuclei are initially only parallel to the beam direction. Transverse field components start to develop later on. We focus here on the longitudinal fields which dominate the glasma's dynamics and are invariant under Lorentz transformations along the beam direction identified with the axis $z$.

\subsection{Field correlation functions}
\label{sec-glasma-corr}

Since the electric and magnetic fields are expressed in the Abelian limit through the four-potential according to Eq.~(\ref{E-B-vs-A-linear}), the potential generating the fields only along the axis $z$ is of the form
\be
\label{A-Ez-Bz}
A^\mu_a(t,{\bf r}) = \big(A^0_a(t,z), A^x_a(x,y), A^y_a(x,y), A^z_a(t,z)\big) ,
\ee
that is the $0$ and $z$ components of $A^\mu$ depend on $t$ and $z$ while the $x$ and $y$ components on $x$ and $y$. The electric field corresponding to the potential (\ref{A-Ez-Bz}) depends on $t$ and $z$ while the magnetic field on $x$ and $y$. The electric field $E^z_a$ is boost invariant if it depends on $t$ and $z$ only through the proper time $\tau \equiv \sqrt{t^2 - z^2}$ which is the Lorentz scalar. However, we do not require the boost invariance of the fields but of the field correlators. To write down a general expression of the electric field correlator $\langle E^z_a (t_1,z_1) \, E^z_b(t_2,z_2) \rangle$, we introduce the variables 
\be
\tau_i \equiv \sqrt{t_i^2 - z_i^2}, ~~~~~~~~~~~~
\eta_i \equiv \frac{1}{2}\log \Big(\frac{t_i + z_i}{t_i - z_i}\Big), ~~~~~~~~~~~~
i = 1,2,
\ee
and we note that the proper times $\tau_i$ and space-time rapidities $\eta_i$ are well defined only for the time-like two-vectors $(t_i,z_i)$. The boost invariant correlation function of the electric fields is assumed to be
\be
\label{corr-EzEz-gen}
\langle E^z_a (t_1,z_1) \, E^z_b(t_2,z_2) \rangle = \delta^{ab}\Theta(t_1^2 - z_1^2) \, \Theta(t_2^2 - z_2^2) \, f_E(\tau_1 - \tau_2,\eta_1 - \eta_2), 
\ee
where $\Theta(t_i^2 - z_i^2) $ is the Heaviside step function and $f_E(\tau ,\eta)$ is an arbitrary function. Since $\tau_i$ and $\eta_i$ are well defined only for $t_i^2 \ge z_i^2$, we require that the correlation function (\ref{corr-EzEz-gen}) vanish when the space-time points  $(t_i,z_i)$ are localized beyond the light cone of the point $(0,0)$. We observe that the correlation function, which depends not only on $\tau_1 - \tau_2$  but on both $\tau_1$ and $\tau_2$ ,  is also boost invariant but we assume the translation invariance in the $\tau$ variable just for simplicity. 

The boost invariant magnetic field correlator is chosen to be
\be
\langle B^z_a (x_1,y_1) \, B^z_b(x_2,y_2) \rangle = \delta^{ab} f_B(x_1 - x_2,y_1-y_2) ,
\ee
that is the plasma is assumed to be translationally invariant in the $x$ and $y$ directions. We note that the fields $E^z_a (t,z)$ and $B^z_a (x,y)$ are completely decoupled from each other in the Abelian limit and thus the mixed correlator $\langle E^z_a (t_1,z_1) \, B^z_b(t_2,z_2) \rangle$ is expected to be small or vanish. 

We introduce the unit vector ${\bf n} = (0,0,1)$ along the axis $z$ and the correlation functions of electric and magnetic fields are written as 
\ba
\label{EzEz-corr-BI}
\langle E^i_a (t_1,{\bf r}_1) \, E^j_b (t_2,{\bf r}_2) \rangle &=& \delta^{ab}
n^i n^j \Theta(t_1^2 - z_1^2) \, \Theta(t_2^2 - z_2^2) \, f_E(\tau_1 - \tau_2,\eta_1 - \eta_2), 
\\[2mm]
\label{BzBz-corr-BI}
\langle B^i_a (t_1,{\bf r}_1) \, B^j_a (t_2,{\bf r}_2) \rangle &=& \delta^{ab} n^i n^j f_B(x_1 - x_2,y_1-y_2) .
\ea
The correlators are chosen to be of the Gaussian form
\ba
\label{EzEz-gauss}
f_E(\tau,\eta)
&=& \tilde{M}_E \exp\Big( - \frac{\tau^2}{2\sigma^2_\tau} - \frac{\eta^2}{2\sigma^2_\eta} \Big) ,
\\[2mm]
\label{BzBz-gauss}
f_B(x,y)
&=&  \tilde{M}_B  \exp\Big( - \frac{x^2+y^2}{2\sigma^2_T} \Big)  ,
\ea
with the real positive parameters $\tilde{M}_E $, $\tilde{M}_B$, $\sigma_\tau$, $\sigma_\eta$ and $\sigma_T$ to be determined later on.

\subsection{Computation of  $X$ and $Y$}

Substituting the correlation function (\ref{BzBz-corr-BI}) with (\ref{BzBz-gauss}) into Eq.~(\ref{X-def}), the magnetic contribution to the tensor $X^{ij}({\bf v})$ is found as
\ba
\label{X-B-1}
X^{ij}_B({\bf v}) = g^2 C_F 
V^{ij} \tilde{M}_B \int_0^t dt' \, \exp\Big( - \frac{(v_x^2 +v_y^2)t'^2}{2\sigma^2_T} \Big) ,
\ea
where 
\be
V^{ij} \equiv \epsilon^{ikl}v^k n^l  \epsilon^{jmn}v^m n^n  = \left( \begin{array}{c c c}
    v_y^2    &  -v_x v_y  & 0 \\
   -v_x v_y  &  v_x^2     & 0 \\
       0        &     0         & 0 \\
\end{array} \right) .
\ee

Since the Fokker-Planck equation is derived in the long time limit, we assume that $t \gg \sigma_T$ and the integral becomes Gaussian. Thus, we get
\ba
\label{X-B-2}
X^{ij}_B({\bf v}) =  \sqrt{\frac{\pi }{2}} \; g^2  C_F \frac{V^{ij}}{v_T} \; \tilde{M}_B \sigma_T ,
\ea
where $v_T \equiv |{\bf v}_T| = \sqrt{v_x^2 +v_y^2}$ is the quark velocity perpendicular to ${\bf n}$. 

The electric contribution to the tensor $X^{ij}({\bf v})$ is
\ba
\label{X-E-1}
X^{ij}_E({\bf v}) &=& g^2 C_F \,
n^i n^j\int_0^t dt' \, \Theta(t^2 - z^2) \, \Theta\Big( t'^2 - \big(z - v_z(t-t') \big)^2\Big)\, f_E (\tau, \eta )  ,
\ea
where
\ba
\label{tau-def}
\tau &\equiv& \tau_1 - \tau_2 = \sqrt{t^2 - z^2} - \sqrt{t'^2 - \big(z - v_z(t-t')\big)^2} ,
\\[2mm]
\label{eta-def}
\eta &\equiv& \eta_1 - \eta_2 = \frac{1}{2}\log \Big(\frac{t + z}{t - z}\Big) 
- \frac{1}{2}\log \bigg(\frac{t'+ \big(z - v_z(t-t')\big)}{t' - \big(z - v_z(t-t')\big)}\bigg) .
\ea

\begin{figure}[t]
\begin{minipage}{87mm}
\centering
\includegraphics[scale=0.24]{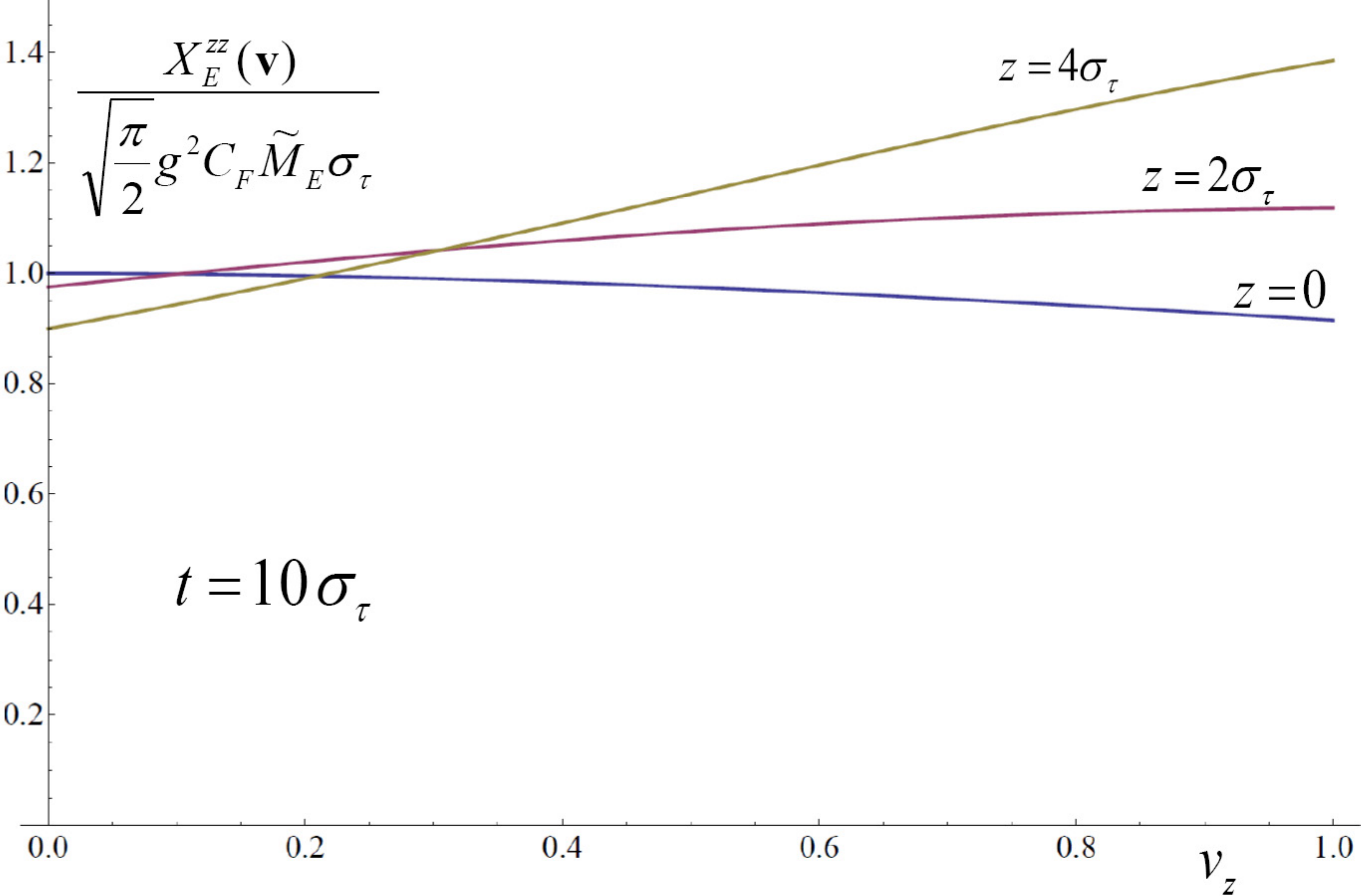}
\end{minipage}
\hspace{1mm}
\begin{minipage}{87mm}
\centering
\includegraphics[scale=0.24]{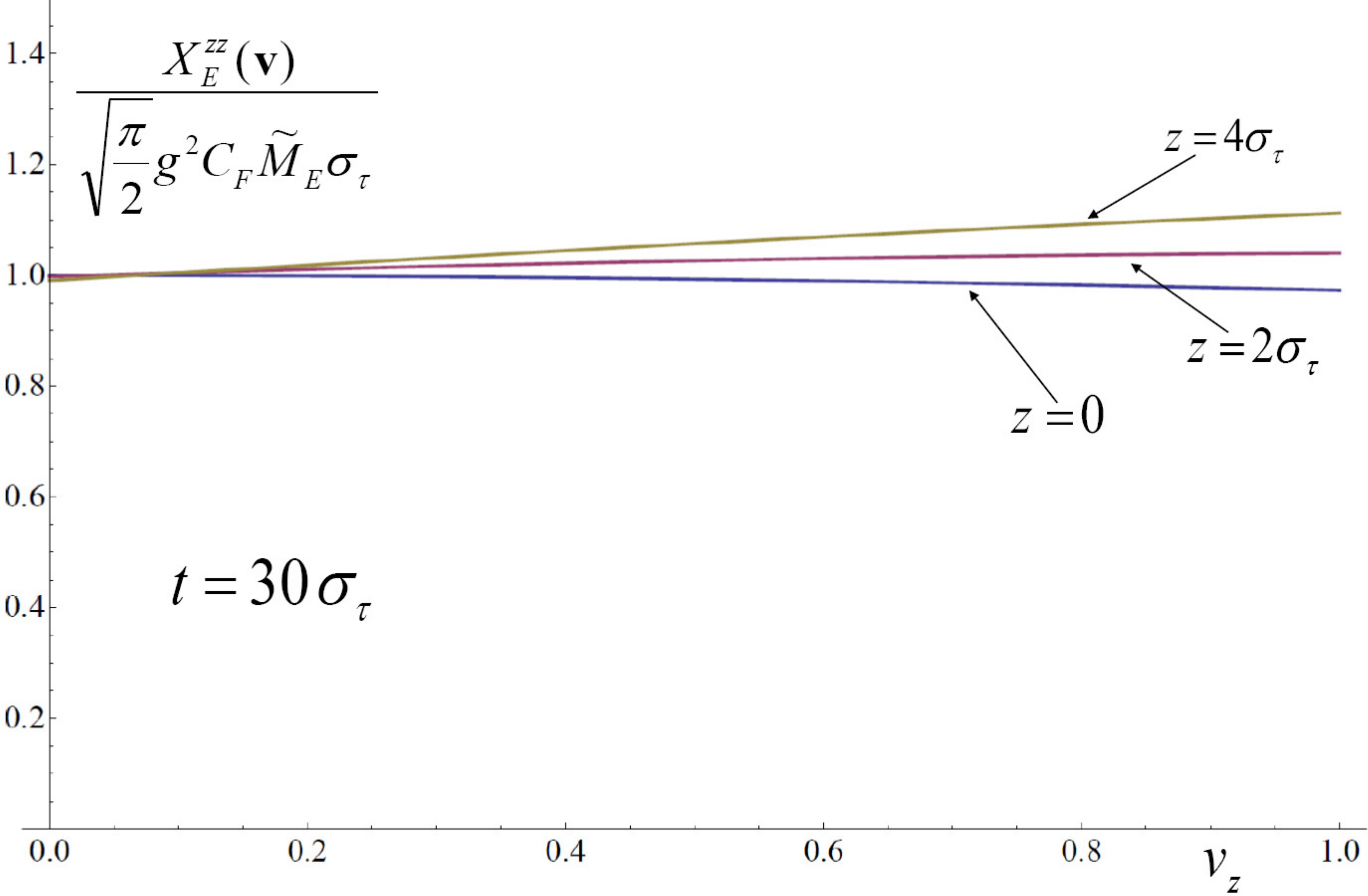}
\end{minipage}
\caption{The coefficient $X^{zz}_E({\bf v})$ divided by the approximate expression (\ref{X-E-approx}) for $t = 10 \sigma_\tau$ in the left panel and for $t = 30 \sigma_\tau$ in the right panel.}
\label{fig-XE-t}
\end{figure}

The integral from Eq.~(\ref{X-E-1}) with the correlation function (\ref{EzEz-gauss}) is difficult to compute analytically. The problem greatly simplifies in the long time limit. When $t \gg z$ and $t \gg \sigma_\tau$ one easily finds 
\ba
\label{X-E-approx}
X^{ij}_E({\bf v}) \approx  \sqrt{\frac{\pi }{2}} \;  g^2 C_F \, n^i n^j \tilde{M}_E \,\sigma_\tau ,
\ea
which is independent of $\sigma_\eta$. We have checked numerically a quality of the approximation (\ref{X-E-approx}). In Fig.~\ref{fig-XE-t} we show the numerically computed coefficient $X^{zz}_E({\bf v})$ divided by the approximate expression (\ref{X-E-approx}) for $t = 10 \sigma_\tau$ and $t = 30 \sigma_\tau$ as a function of heavy-quark velocity $v_z$.  The coefficient is independent of $v_x$ and $v_y$. The correlation length in rapidity is chosen as $\sigma_\eta=1$ but for a bigger $\sigma_\eta$ the approximation (\ref{X-E-approx}) is even more accurate. We conclude that the approximation (\ref{X-E-approx}) works pretty well and 
we write the tensor $X^{ij}({\bf v})$, which includes the magnetic and electric contributions, as
\be
\label{X-final}
X^{ij}({\bf v}) =  \sqrt{\frac{\pi }{2}} \; g^2 C_F 
\bigg( n^i n^j \tilde{M}_E \,\sigma_\tau + \frac{V^{ij}}{v_T} \; \tilde{M}_B \sigma_T\bigg) .
\ee

Using the relation (\ref{eq-X-vs-Y}), one finds that the vector $ Y^i({\bf v})$ equals
\be
\label{Y-formula}
 Y^i({\bf v}) = \sqrt{\frac{\pi }{2}} \; g^2 C_F 
({\bf v} \cdot {\bf n}) \, n^i  \frac{\tilde{M}_E \sigma_\tau }{T}  .
\ee
The temperature will be estimated in the next section but, as already discussed at the beginning of Sec.~\ref{sec-turbulent-QGP}, the result (\ref{Y-formula}) should be treated with a reservation. 

\subsection{Estimates of parameters}
\label{sec-estimates}

To get approximate values of the parameters, which characterize the glasma, let us estimate the density of energy released in relativistic heavy-ion collisions. When one deals with the central collisions of two nuclei of mass number $A$, the energy density in the center-of-mass frame of colliding nuclei is roughly
\be
\varepsilon_{\rm coll} = \frac{c_{\rm inel} A \sqrt{s}}{\pi R_A^2 l} ,
\ee
where $\sqrt{s}$ is the center-of-mass energy of nucleon-nucleon collision, $c_{\rm inel}$ is the inelasticity parameter which gives the fraction of all accessible energy going to particle production, $R_A$ is the radius of colliding nuclei and $l$ is a length of the cylinder where the energy is released. Assuming that  $c_{\rm inel}=0.5$  \cite{Navarra:2003am} and taking $A=200$, $R_A = 7$ fm and $l = 1$ fm, one obtains $\varepsilon_{\rm coll} \approx 3.25 \; {\rm TeV/fm^3}$ for $\sqrt{s}$ = 5 TeV which is the energy of Pb-Pb collisions at LHC of the 2015 run. 

According to Eq.~(\ref{en-den-field}) the density of energy accumulated in chromodynamic fields equals  
\ba
\label{energy-den-fields}
\varepsilon_{\rm field} = \frac{N_c^2 -1}{2} \big(f_E(0,0) + f_B(0,0) \big) 
= \frac{N_c^2 -1}{2}  (\tilde{M}_E + \tilde{M}_B) ,
\ea
where the explicit form of the correlation functions  (\ref{EzEz-gauss}) and  (\ref{BzBz-gauss}) have been used. The field energy density is controlled by the parameters $\tilde{M}_E$ and $\tilde{M}_B$ but is independent of the correlations lengths $\sigma_T$, $\sigma_\tau$ and $\sigma_\eta$. Assuming $\tilde{M}_E = \tilde{M}_B$, as suggested \cite{Chen:2015wia}, we get $\varepsilon_{\rm field} = 8 \tilde{M}_E$ for $N_c=3$. Requiring that $\varepsilon_{\rm field} = \varepsilon_{\rm coll}$, the parameter $\tilde{M}_E$ is estimated as 
\be
\tilde{M}_E= \frac{1}{8} \, \varepsilon_{\rm coll} \approx 3.10~{\rm GeV}^4.
\ee

To use the formula (\ref{eq-X-vs-Y}) to obtain the coefficient $Y^i({\bf v})$, one needs a temperature of the equilibrated quark-gluon plasma of the same energy density as the glasma. Using the formula (\ref{en-den-QGP}), the equation $\varepsilon_{\rm field} = \varepsilon_{\rm QGP}$ provides 
\be
T = \Big(\frac{240}{37 \pi^2} \, \tilde{M}_E \Big)^{1/4} \approx 1.20~{\rm GeV} .
\ee
where $N_c =3$ and $N_f =2$.

Within the CGC approach the strong longitudinal chromodynamic fields are screened on transverse distance which is of the order of the inverse saturation momentum $Q_s$. Since $Q_s$ is estimated as 2 GeV \cite{Gelis:2012ri}, we choose the transverse correlation $\sigma_T = Q_s^{-1} = 0.5~{\rm GeV}^{-1}$. We also assume that  $\sigma_\tau = \sigma_T$. The remaining parameter is the coupling constant which, as previously, is chosen to be $\alpha_s =0.1$. As already mention in Sec.~\ref{sec-eq-vs-turb}, the glasma magnetic field is stable against the Nielsen-Olesen instability, because the inhomogeneity length of order $Q_s^{-1}$ is not much bigger than the Larmor radius $(gB)^{-1/2}$, $Q_s$ is even bigger than $\tilde{M}_B^{1/4}$. 

Using the formulas (\ref{Y-formula}) and (\ref{X-final}) combined with Eqs.~(\ref{eloss-Y}) and (\ref{qhat-X-T}), we obtain the following estimates of the energy loss and momentum broadening 
\ba
\label{Eloss-num}
- \frac{dE}{dx} &=& \frac{2^{7/2} \pi^{3/2}}{3} \; \alpha_s v \cos^2\!\theta \,  \frac{\tilde{M}_E \sigma_\tau}{T}
\approx 14 ~ v \cos^2\!\theta ~~ \Big[\frac{\rm GeV}{\rm fm} \Big],
\\[2mm]
\label{qhat-num}
\hat{q} &=& \frac{2^{9/2} \pi^{3/2}}{3} \; \alpha_s
\frac{\sin^2\!\theta  + v \sin\theta}{v} \; \tilde{M}_E \sigma_\tau
\approx 33 ~ \frac{\sin^2\!\theta  + v \sin\theta}{v} ~~ \Big[\frac{{\rm GeV}^2}{\rm fm} \Big],
\ea
where $\theta$ is the angle between ${\bf v}$ and ${\bf n}$. Because the electric field is along the beam axis, the collisional energy loss is maximal when the heavy quark moves along the axis, ${\bf v} \parallel {\bf n}$. The maximal momentum broadening occurs when ${\bf v} \perp {\bf n}$. 

Let us compare the numerical values (\ref{Eloss-num}) and (\ref{qhat-num}) with those which are required to properly model experimental data on the charm meson suppression. The collisional energy loss and momentum broadening of a charm quark with 10 GeV momentum in the plasma of the temperature from the interval 0.35 -- 0.5 GeV are estimated \cite{Prino:2016cni} as 
\ba
\label{Eloss-exp}
- \frac{dE}{dx} &=&  1.0\!-\!3.0  ~~ \Big[\frac{\rm GeV}{\rm fm} \Big],
\\[2mm]
\label{qhat-exp}
\hat{q} &=& 1.5\!-\!7.0 ~~ \Big[\frac{{\rm GeV}^2}{\rm fm} \Big]. 
\ea
As seen, the values (\ref{Eloss-num}) and (\ref{qhat-num}) can be significantly larger than (\ref{Eloss-exp}) and (\ref{qhat-exp}), suggesting that in spite of a short lifetime of the glasma it can provide a significant contribution to the collisional and radiative energy loss to heavy quarks from relativistic heavy-ion collisions. Consequently, the effect should be included in the phenomenology of jet quenching.

\section{Summary, Conclusions and Outlook}
\label{sec-conclusions}

Applying the so-called quasi-linear theory \cite{Ved61,LP81}, we have derived a general form of the Fokker-Planck equation of heavy quarks embedded in the plasma of light quarks and gluons. Since the interaction is taken into account through the correlation functions of chromodynamic fields, heavy quarks are seen as interacting not with plasma constituents but rather with fields present in the plasma. At first we have obtained the explicit form of the equation for the case of equilibrium plasma which was studied long ago \cite{Svetitsky:1987gq} using the standard method where the Fokker-Planck equation simply approximates the Boltzmann one. Although our approach is noticeably different, the Fokker-Planck equation we obtained agrees with the standard one \cite{Svetitsky:1987gq}. 

In the second part of the paper, the method developed for the equilibrium plasma has been applied to the turbulent plasma populated with strong fields. The parametric estimate shows that the interaction of heavy quarks with the turbulent plasma is much stronger than with the equilibrium one of the same energy density if the coupling constant is truly small. The effect is less prominent  for a realistic value of the coupling constant and the difference depends on characteristics of the plasma fields. Within the `Gaussian E \& B' model both the energy loss and momentum broadening are significantly bigger than the equilibrium results. A dependence of $dE/dx$ and $\hat{q}$ on heavy quark velocity also strongly depends on how the turbulent plasma is modeled.  

The third part of our study is devoted to the glasma from the earliest stage of relativistic heavy-ion collisions. Assuming that there are chromoelectric and chromomagnetic fields only along the beam direction we have derived the appropriate Fokker-Planc equation. We have also shown that in spite of its short lifetime the glasma can provide a significant contribution to the collisional and radiative energy loss of heavy quarks. 

Our findings clearly suggest a direction of further work. We need a more realistic model of turbulent QCD plasma from relativistic heavy-ion collisions. In contrast to the simple model discussed here, a temporal evolution of the glasma has to be taken into account and the fields cannot be purely longitudinal. The CGC studies \cite{Gelis:2012ri} and, in particular, the analytic analysis \cite{Chen:2015wia} provide a very good guidance to build up such a model.  

\section*{Acknowledgments}

I am very grateful to Margaret Carrington for numerous fruitful discussions and valuable comments.

\appendix

\section{Equilibrium correlation functions}
\label{sec-corr-fun}

The correlation functions of chromodynamic fields in the equilibrium plasma, which were studied in detail in \cite{Mrowczynski:2008ae}, can be expressed as 
\ba
\label{EE-1}
\langle H^i_a(t,{\bf r}) K^j_b (t',{\bf r}')\rangle &=&
\int {d\omega \over 2\pi} \int {d^3k \over (2\pi)^3}
e^{-i\big(\omega (t-t') - {\bf k} \cdot ({\bf r} - {\bf r}')\big)} \langle H^i_a K^j_b \rangle_{\omega,\, {\bf k}} ,
\ea
where $H^i_a(t,{\bf r})$ and $K^j_b(t,{\bf r})$ is either the electric or magnetic field, and the fluctuation spectra are
\ba
\label{EiEj-spec-eq}
\langle E^i_a E^j_b\rangle_{\omega,\, {\bf k}}
&=& 
2 \delta^{ab} \frac{\omega^4}{e^{\beta  |\omega|} - 1}  \bigg[
\frac{k^ik^j}{{\bf k}^2}
\frac{\Im \varepsilon_L(\omega,{\bf k})}
{|\omega^2 \varepsilon_L(\omega,{\bf k})|^2}
+ \Big(\delta^{ij} - \frac{k^ik^j}{{\bf k}^2}\Big)
\frac{\Im \varepsilon_T(\omega,{\bf k})}
{|\omega^2 \varepsilon_T(\omega,{\bf k})-{\bf k}^2|^2}
\bigg] ,
\\[2mm]
\label{BiBj-spec-eq}
\langle B^i_a B^j_b\rangle_{\omega,\, {\bf k}}
&=& 2 \delta^{ab}  \frac{\omega^2 {\bf k}^2}{e^{\beta  |\omega|} - 1} 
 \Big(\delta^{ij} - \frac{k^i k^j}{{\bf k}^2} \Big) 
\frac{\Im \varepsilon_T (\omega,{\bf k})}
{|\omega^2 \varepsilon_T(\omega,{\bf k})-{\bf k}^2|^2} ,
\\[2mm]
\label{BiEj-spec-eq}
\langle B^i_a E^j_b\rangle_{\omega,\, {\bf k}}
&=& \langle E^j_a B^i_b\rangle_k
=2 \delta^{ab}  \frac{\omega^3}{e^{\beta |\omega|} - 1} \epsilon^{imj}k^m
\frac{\Im \varepsilon_T (\omega,{\bf k})}
{|\omega^2 \varepsilon_T(\omega,{\bf k})-{\bf k}^2|^2} ,
\ea
$\epsilon^{ijm}$ is the antisymmetric tensor, $\beta \equiv T^{-1}$ with $T $ being the system's temperature and $\varepsilon_{L,T}(\omega,{\bf k})$ are chromodielectric functions. For an equilibrium plasma of massless particles, the functions
are well-known to be, see {\it e.g.} \cite{lebellac},
\ba
\label{Re-eL-massless}
\Re\varepsilon_L(\omega,{\bf k}) 
&=& 
1+ \frac{m_D^2}{{\bf k}^2}
\bigg[
1 - \frac{\omega}{2|{\bf k}|}
{\rm ln}\bigg|\frac{\omega + |{\bf k}|}{\omega - |{\bf k}|} \bigg| 
\bigg] , 
~~~~~~~~~~~~~~
\Im\varepsilon_L(\omega,{\bf k}) 
= \frac{\pi}{2} \: 
\Theta ({\bf k}^2 -\omega^2) \:
\frac{m_D^2 \omega}{|{\bf k}|^3} ,
\\[4mm]
\label{Re-eT-massless}
\Re\varepsilon_T(\omega,{\bf k}) 
&=& 
1 - \frac{m_D^2}{2{\bf k}^2}\bigg[
1 - \frac{\omega^2 - {\bf k}^2}{2\omega |{\bf k}|}
{\rm ln}\bigg|\frac{\omega + |{\bf k}|}{\omega - |{\bf k}|} \bigg|
\bigg] , 
~~~~~~~~~
\Im\varepsilon_T(\omega,{\bf k}) 
= \frac{\pi}{4} \:
\Theta ({\bf k}^2 -\omega^2) \:
\frac{m_D^2({\bf k}^2- \omega^2)}{\omega |{\bf k}|^3} ,
\ea
with $m_D$ being the Debye mass which for the quark-gluon plasma of $N_f$ massless flavors equals 
\be
\label{m_D}
m_D^2 = \frac{g^2 T^2}{6}\, (N_f + 2N_c) .
\ee
When $\omega^2 \ll {\bf k}^2$, the dielectric functions can be approximated as
\ba
\label{Re-Im-eL-small-omega}
\Re\varepsilon_L(\omega,{\bf k}) 
&=& 
1+ \frac{m_D^2}{{\bf k}^2} , 
~~~~~~~~~~~~~~~~~~~~
\Im\varepsilon_L(\omega,{\bf k}) 
= \frac{\pi}{2} \: 
\frac{m_D^2 \omega}{|{\bf k}|^3} ,
\\[2mm]
\label{Re-Im-eT-small-omega}
\Re\varepsilon_T(\omega,{\bf k}) 
&=& 
1 - \frac{m_D^2}{{\bf k}^2} , 
~~~~~~~~~~~~~~~~~~~~
\Im\varepsilon_T(\omega,{\bf k}) 
= \frac{\pi}{4} \:
\frac{m_D^2}{\omega |{\bf k}|} .
\ea

We note that the fluctuation spectra of pure classical fields were actually derived in our study \cite{Mrowczynski:2008ae}. The effect of Bose statistics of field quanta has been included in the formulas (\ref{EiEj-spec-eq}), (\ref{BiBj-spec-eq}), and (\ref{BiEj-spec-eq}) by means of the substitution
\be
\label{replace-quant}
\frac{T}{\omega} \to \frac{\rm{sgn}(\omega)}{e^{\beta |\omega|} - 1} ~ \stackrel{\omega \ll T}{\approx} ~ \frac{T}{\omega}.
\ee
The absolute value of the frequency results from the following reasoning. Since the electric and magnetic fields are real in coordinate space, their correlation functions are real as well. Consequently, the fluctuation spectra must obey
\be
\label{mirror-sym}
\langle H^i_a K^j_b\rangle_{\omega,\, {\bf k}} = \langle H^i_a K^j_b\rangle_{-\omega,\, -{\bf k}} .
\ee 
One checks that the formulas (\ref{EiEj-spec-eq}), (\ref{BiBj-spec-eq}), and (\ref{BiEj-spec-eq}) indeed satisfy the symmetry (\ref{mirror-sym}) and thus the correlation function (\ref{EE-1}) is real as it should be.

\section{Correlation function and fluctuation spectrum}
\label{sec-corr-spec}

We discuss here the relation between the correlation function of the Fourier transformed fields and the fluctuation spectrum. The correlation function of the Fourier transformed fields equals
\be
\langle H^i_a(\omega, {\bf k}) \, K^j_b(\omega', {\bf k}') \rangle 
= \int dt \,d^3r \, e^{i(\omega t - {\bf k}\cdot {\bf r})}
\int dt' \, d^3r' e^{i(\omega' t' - {\bf k}'\cdot {\bf r}')}
\langle H^i_a(t, {\bf r}) \, K^j_b(t', {\bf r}') \rangle ,
\ee
where $H^i_a(t,{\bf r})$ and $K^j_b(t,{\bf r})$ is either the electric field, magnetic field, or the potential. To define the fluctuation spectrum we first write down the correlation function as
\be
\langle H^i_a(t_1, {\bf r}_1) \, K^j_b(t_2, {\bf r}_2) \rangle = 
\langle H^i_a \big(t+\frac{\Delta t}{2}, {\bf r} + \frac{\Delta{\bf r}}{2} \big) \, 
K^j_b\big(t - \frac{\Delta t}{2}, {\bf r} - \frac{\Delta{\bf r}}{2}\big) \rangle , 
\ee
where the new space-time variables read
\ba
t &\equiv& \frac{t_1+t_2}{2},
~~~~~~~~~~~~~~~~~~{\bf r} \equiv \frac{{\bf r}_1 + {\bf r}_2}{2} ,
\\
\Delta t &\equiv& t_1 - t_2 ,
~~~~~~~~~~~~~~~~~~~~\Delta {\bf r} \equiv {\bf r}_1 - {\bf r}_2 .
\ea
The fluctuation spectrum is defined as
\ba
\label{spec-RT}
\langle H^i_a K^j_b \rangle_{\omega, {\bf k}} = \int d\Delta t \,d^3\!\Delta r \, 
e^{i(\omega \Delta t - {\bf k}\cdot \Delta {\bf r})}
\langle H^i_a \big(t+\frac{\Delta t}{2}, {\bf r} + \frac{\Delta{\bf r}}{2} \big) \, 
K^j_b\big(t - \frac{\Delta t}{2}, {\bf r} - \frac{\Delta{\bf r}}{2}\big) \rangle  .
\ea
In general, the fluctuation spectrum (\ref{spec-RT}) depends on $t$ and ${\bf r}$. However, if the system is stationary and homogeneous, that is translationally invariant in both space and time, the spectrum is independent of $t$ and ${\bf r}$. Therefore, Eq.~(\ref{spec-RT}) can be rewritten as
\ba
\label{spec-RT-int}
\langle H^i_a K^j_b \rangle_{\omega, {\bf k}} =  
\frac{1}{V {\cal T}}\int dt \,d^3r \int d\Delta t \,d^3\!\Delta r \, 
e^{i(\omega \Delta t - {\bf k}\cdot \Delta {\bf r})}
\langle H^i_a \big(t+\frac{\Delta t}{2}, {\bf r} + \frac{\Delta{\bf r}}{2} \big) \, 
K^j_b\big(t - \frac{\Delta t}{2}, {\bf r} - \frac{\Delta{\bf r}}{2}\big) \rangle  ,
\ea
where $V{\cal T}$ is the space-time volume occupied by the system. Substituting the space-time correlation function expressed through $\langle H^i_a(\omega, {\bf k}) \, K^j_b(\omega', {\bf k}') \rangle $ into Eq.~(\ref{spec-RT-int}) and performing the trivial integrations involving delta functions, one obtains the desired relation
\be
\label{fluc-spec-vs-FT}
\langle H^i_a K^j_b \rangle_{\omega, {\bf k}} = \frac{1}{V {\cal T}} \, 
\langle H^i_a(\omega, {\bf k}) \, K^j_b(-\omega, -{\bf k}) \rangle .
\ee


\end{document}